\documentclass[preprint]{ptephy_v1}

\preprintnumber{XXXX-XXXX} 
\usepackage{hyperref}

\usepackage{longtable} 
\usepackage{caption}
\begin{document}

\title{Nuclear masses and the equation of state of nuclear matter}


\author{Kazuhiro Oyamatsu}
\affil{Department of Human Informatics, Aichi Shukutou University, 2-9 Katahira, Nagakute , Aichi 480-1197, Japan \email{oyak@asu.aasa.ac.jp}}

\begin{abstract}%
The incompressible liquid-drop (ILD) model reproduces masses of stable nuclei rather well. 
Here we show how the ILD volume, surface, symmetry,  and Coulomb energies are related to the equation of state of nuclear matter using the Oyamatsu-Iida (OI) macroscopic nuclear model, which has reasonable many-body energy and isoscalar inhomogeneity gradient energy. 
We use 304 update interactions, covering wide ranges of the incompressibility $K_0$ of symmetric matter and the density slope of symmetry energy $L$,  which fit almost equally empirical mass and radius data of stable nuclei.  Thus, the $K_0$ and $L$ dependences are nearly frozen in stable nuclei as in the ILD model, leading to clear correlations among interaction and saturation parameters. 
Furthermore, we assume that the surface energy of the OI model is twice as large as the gradient energy using the size equilibrium conditions of the ILD and OI models. Then, the four energies of the ILD and OI models agree well for stable nuclei with $A \gtrsim 40$. 
Meanwhile, the OI model with $L \lesssim 100$ MeV predicts the latest mass data better than those of stable nuclei,  and we suggest $20 \lesssim  L \lesssim 90$ MeV, although the lower boundary is not constrained well.
\end{abstract}

\subjectindex{xxxx, xxx}

\maketitle

\section{Introduction}
The incompressible liquid drop (ILD) model, also referred to as the Weizs\"acker-Bethe mass formula,  assumes the same sharp nuclear surface for the neutron and proton distributions and fits rather well the observed masses and neutron excesses of the $\beta$-stable nuclei \cite{BohrMottelson1968}.
Meanwhile, the observed radii can also be reproduced well in the macroscopic nuclear model using the equation of state (EOS) of uniform nuclear matter and appropriate inhomogeneity energy correction due to finite-range effects of nuclear forces \cite{Lombard1973}. 


The Oyamatsu-Iida (OI) macroscopic nuclear model \cite{OI2003} for studies on laboratory nuclei and neutron star matter \cite{OI2007,Oyamatsu:2010bf,Oyamatsu:2010sk,Iida:2013fra,Sotani:2013dga,Sotani:2012qc,Sotani:2012xd,Sotani:2013jya, Sotani:2015lya, Sotani:2015laa, Sotani:2016pmb,Sotani:2017hpq,Sotani:2018tdr, Sotani:2019pja} is one of the latter type models. It is essentially based on the assumption of the density functional theory (DFT), initially developed for interacting electrons, which states the total energy can be written as a functional of the local density  \cite{Hohenberg:1964zz, Kohn:1965zza}. 
The introduction section of Ref. \cite{Lombard1973} gives a concise discussion of macroscopic nuclear models at the dawn of the DFT.

The OI model has three distinct features from other macroscopic models using phenomenological nuclear interactions, such as the Skyrme Hartree-Fock theory and relativistic mean field theory (see, for example, Ref. \cite{BrackGuetHakansson1985} for a review of microscopic and macroscopic descriptions using Skyrme interactions). 
First, the OI model parameterizes the EOS and inhomogeneity energy directly, although it loses the direct connection between nuclear forces and the EOS. 
Special attention is paid to the incompressibility $K_0$ and the slope of symmetry energy $L$ of the EOS because the nuclear structure is determined from the local pressure equilibrium. 
Second, the OI model is designed as a compressible liquid drop model allowing the independent radius and surface diffuseness parameters for the neutron and proton distributions.
Third, to smooth out the shell effects, the interaction parameters of the OI model are fitted to the smoothed empirical mass and radius data of stable nuclei  rather than the experimental data. These smoothed data \cite{MYamada1964} were evaluated from the systematics \cite{YamadaMatumoto1961JPSJ}, and eventually, the neutron excess values take decimal values rather than integers.



The OI model has been used to study unstable nuclei and neutron star matter \cite{OI2007,Oyamatsu:2010bf,Oyamatsu:2010sk,Iida:2013fra,Sotani:2013dga,Sotani:2012qc,Sotani:2012xd,Sotani:2013jya, Sotani:2015lya, Sotani:2015laa, Sotani:2016pmb,Sotani:2017hpq,Sotani:2018tdr, Sotani:2019pja}.
Therefore, it is time to summarize the results of these studies and clarify how the EOS of nuclear matter affects the structures of the laboratory nuclei and neutron star matter. In this paper, we reexamine how many EOS saturation parameters are constrained by the empirical mass and radius data of stable nuclei. Then, we show that masses of unstable nuclei correlate with the $L$ value and compare this correlation with the recently reported $L$ values evaluated from $^{208}$Pb neutron skin and Sn+Sn experiments \cite{Reed:2021nqk, Reinhard:2021utv, SRIT:2021gcy}.
Furthermore, we discuss how the EOS and the inhomogeneity energy are related to the ILD volume, surface, symmetry, and Coulomb energies, showing that the OI model has two more degrees of freedom of $K_0$ and $L$ than the ILD model. 
We will also take a look at the surface diffuseness in the most stable nuclei because the nuclear density distributions are not yet satisfactory in the macroscopic nuclear model \cite{Lombard1973}.
In future papers, we plan to discuss the neutron drip and neutron star matter based on the results of this paper.

This paper is arranged as follows. Section \ref{macroscopic_nuclear_model} describes the OI model in detail.   Section  \ref{uniform_matter_energy_EOS_saturation_patrameters} defines uniform matter energy, density-dependent symmetry energy, and saturation parameters. Section \ref{nucleus_in_the_OI_model} describes a nucleus in the OI model. In Sec. \ref{2017OIEOS_optimization}, we update the values of the interaction parameters and show to what extent the updated interactions fit the empirical data of stable nuclei.
Section  \ref{OIEOSF2017_interaction_saturation_paramter_values}  shows the obtained correlations among the interaction and saturation parameters and gives the numerical results of nuclear mass calculations. 
Section \ref{LiquidDrop_energies_in_OI_model} shows how the volume, surface, and symmetry energies of the ILD model are represented in the OI model.   Finally, the conclusions of this paper are given in Sec. \ref{conclusions}.



\section{Oyamatsu-Iida macroscopic nuclear model}

\label{macroscopic_nuclear_model}

The Oyamatsu-Iida (OI) macroscopic nuclear model \cite{OI2003} is an update of model IV for the early study of pasta nuclei in the neutron-star crust \cite{oya1993}. The OI model  
has the following three important features compared to the ILD model.

\begin{itemize}
\item Nuclear energy in a nucleus is the integral of the local uniform-matter and inhomogeneity energy densities.
\item The inhomogeneity energy density is proportional to the square of the gradient of the local nucleon density.
\item The neutron and proton distributions are independent; each distribution is parameterized with radius and diffuseness parameters.
\end{itemize}

\subsection{Uniform-matter energy density  $\epsilon_0 \left(n_n,n_p\right)$}
\label{uniform_matter_energy_EOS_saturation_patrameters}

We write the energy density $\epsilon_0 \left(n_n,n_p\right)$ of uniform nuclear matter as the sum of the free kinetic energy density and the potential energy density. The potential energy density is  the weighted sum of $v_s(n)$ for symmetric matter and $v_n(n)$ for neutron matter.
\begin{equation}
\epsilon_0 \left(n_n,n_p\right) = \frac{3}{5} (3 \pi^2)^{2/3} \left(\frac{\hbar^2}{2m_n} n_n^{5/3}+\frac{\hbar^2}{2m_p} n_p^{5/3}\right) + (1 -\alpha^2) v_s(n) +\alpha^2 v_n(n)
\label{homogineoues_energy_density}
\end{equation}
with $\alpha=(n_n-n_p)/n$.
These potential energy densities are parametrized as
\begin{equation}
v_s(n)=a_1 n^2 + \frac{a_2 n^3}{1+a_3 n}, \quad v_n(n)=b_1 n^2 + \frac{b_2 n^3}{1+b_3 n} .
\label{BuldmanDoverPotential}
\end{equation}
The coefficients $a_1 \ (b_1)$ and $a_2 \ (b_2)$ are two- and three-body energy coefficients for symmetric (neutron) matter, respectively. 
The coefficients $a_3$ and  $b_3$ are the many-body parameters that control the strength of many-body ($N \ge 4$) energies. 
For example, the potential energy density $v_s(n)$ can be expanded as
\begin{equation}
v_s(n)=a_1 n^2 + a_2 n^3 [1-a_3 n+ (a_3 n)^2 - (a_3 n)^3 + \cdots].
\label{PotentialExpansion}
\end{equation}
Here, the two-body, three-body, and $N-$body  ($N \ge 4$) energy densities are $a_1 n^2$, $a_2 n^3$, and  $a_2 n^3 (-a_3 n)^{N-3}$, respectively.
The potential energy density of the form in Eq. (\ref{BuldmanDoverPotential}) 
was proposed by Buldman and Dover\cite{BludmanDover1980} to make the equation of state soft and causal at high densities. It can fit the popular nuclear matter EOS by Friedman and Pandharipande (FP) \cite{FriedmanPandharipande1980} up to $n=0.3$ (fm$^{-3}$) \cite{oya1993}.
 However, it is challenging to constrain the many-body parameter $b_3$ of neutron matter from stable nuclei. Therefore, we set $b_3$ = 1.58632 fm$^3$ \cite{oya1993,OI2003}, chosen to fit the FP neutron matter EOS \cite{FriedmanPandharipande1980} to give reasonable many-body energy for neutron matter. As a side note, $v_n(n)$  in the early study\cite{oya1993} has an additional constraint $b_1/b_2=-0.3232$ to fit the FP neutron matter EOS better.

It is convenient to consider the energy per nucleon of the matter as a function of the total nucleon density $n$ and the neutron excess $\alpha=(n_n-n_p)/n$.
This energy per nucleon
\begin{equation}
w(n,\alpha)=\epsilon_0(n_n, n_p)/n
\label{EOS_def}
\end{equation}
is often referred to as the equation of state (EOS). 

Saturation parameters are essentially the density derivative coefficients of $w(n, \alpha)$ at the saturation ($n=n_0$ and $ \alpha=0$);  thereby,  the behavior of the EOS close to the saturation is determined mainly by low order saturation parameters. 
We write the energies of symmetric nuclear matter ($\alpha=0$) and neutron matter ($\alpha=1$) as $w_s(n)=w(n,0)$ and $w_n(n)=w(n,1)$, respectively.
Due to the  charge symmetry property of the nuclear interaction, $w(n, \alpha)$ can be expanded into the Taylor series with respect to $\alpha^2$:
\begin{equation}
w(n,\alpha)=w_s(n) + S^{(2)}(n) \alpha^{2} + \frac{1}{2} S^{(4)}(n) \alpha^{4} + \frac{1}{6} S^{(6)}(n) \alpha^{6} + \cdots ,
\end{equation}
with
\begin{equation}
S^{(2k)}(n)= \frac{\partial^k w}{\partial (\alpha^{2})^k}\Big|_{\alpha=0} \quad(k=1,2, \cdots).
\label{density_dependent_S}
\end{equation}
The energy $S^{(2)}(n)$  dominates the asymmetry energy and is usually referred to as the density-dependent symmetry energy  $S(n)$.

It is useful to expand the three energies $w_s(n)$,  $w_n(n)$, and $S(n)$ in the neighborhood of the saturation density $n=n_0$ using a dimensionless parameter $\displaystyle u=\frac{n-n_0}{3 n_0}$ instead of $n$. 

\begin{equation}
w_s(n)=w_0+L_0 u + \frac{1}{2}K_0 u^2 + \frac{1}{6}Q_0 u^3 + \cdots,
\label{wsn_def}
\end{equation}
\begin{equation}
w_n(n)=w_{n0}+L_{n0} u + \frac{1}{2}K_{n0} u^2 + \frac{1}{6}Q_{n0} u^3 + \cdots,
\label{wnn_def}
\end{equation}
\begin{equation}
S(n)=S_{0}+L u + \frac{1}{2}K_{sym} u^2 + \frac{1}{6}Q_{sym} u^3 + \cdots,
\label{S0n_def}
\end{equation}
The coefficients in Eqs. (\ref{wsn_def})-(\ref{S0n_def}) are called saturation parameters. 
In Eq. (\ref{wsn_def}), the density slope $L_0$ is zero  from the saturation condition. 

This paper only discusses the saturation parameters up to $Q_0$, $Q_{n0}$, and $Q_{sym}$.  These $Q$s  in Eqs. (\ref{wsn_def})-(\ref{S0n_def})  are proportional to the third-order derivative coefficients of density and depend only on the three-body and many-body parameters ($a_2, a_3, b_2$, and $b_3$) in the OI model.
The explicit formula giving relations between the potential parameters ($a_1-a_3, \ b_1-b_3$) and the low order saturation parameters are given in Appendix \ref{explicit_formula_of_saturation_parameters_w_potential_parameters}.

In addition to the saturation parameters in Eqs. (\ref{wsn_def})-(\ref{S0n_def}), we introduce auxiliary parameter $y$, the density slope of the saturation curve at $\alpha=0$, to reasonably constrain isovector interaction.
In the lowest-order approximation,
\begin{equation}
y=-\frac{S_0 K_0}{3 n_0 L}.
\label{slope_y}
\end{equation}

\subsection{Nucleus described in the OI model}
\label{nucleus_in_the_OI_model}

The mass excess, $M_{ex}$, of a charge-neutral atomic nucleus of proton number $Z$, neutron number $N$, and mass number $A=N+Z$ is the sum of the EOS (uniform-matter) energy $W_{EOS}$, the gradient (inhomogeneity) energy $W_g$, the Coulomb energy $W_C$, and the rest mass energy $\Delta m$.
\begin{equation}
M_{ex}  = W_{EOS} + W_g + W_c + \Delta m, 
\label{Mex_def}
\end{equation}
\begin{equation}
\Delta m = m_n N +(m_p + m_e)Z - m_u  A
\label{dm_def}
\end{equation}
with the neutron mass $m_n$, the proton mass $m_p$, the electron mass $m_e$, and the atomic mass unit $m_u$. 
In Eq. (\ref{dm_def}), we use $m_p + m_e$ instead of the hydrogen mass $m_H$. The difference between $m_p + m_e$ and $m_H$ is numerically minor, less than one keV. This use of the electron rest mass $m_e$ is helpful in the neutron-star matter calculation because the electron energy of the neutron-star matter is approximated well by the relativistic electron kinetic energy \cite{oya1993,OI2007}.

The neutron number $N$ (proton number $Z$)  is given by the integral of local neutron  (proton) number density $n_n(r)$ ($n_p(r)$).
\begin{equation}
N=\int d^3 r \  n_n(r),
\label{N_def}
\end{equation}
\begin{equation}
Z=\int d^3 r \  n_p(r).
\label{Z_def}
\end{equation}
The mass number $A$ is given by 
\begin{equation}
A=\int d^3 r \  (n_n(r)+n_p(r)) =\int d^3 r \  n(r)
\label{A_def}
\end{equation}
with the total nucleon number density $n(r)=n_n(r)+n_p(r)$.

As in our previous studies \cite{OI2003,OI2007,Oyamatsu:2010bf,Oyamatsu:2010sk,Iida:2013fra,Sotani:2013dga,Sotani:2012qc,Sotani:2012xd,Sotani:2013jya, Sotani:2015lya, Sotani:2015laa, Sotani:2016pmb,Sotani:2017hpq,Sotani:2018tdr, Sotani:2019pja},
we assume that the local nuclear energy density is the sum of the uniform-matter energy density $\epsilon_0(n_n,n_p)$ and the gradient energy density $F_0 |\nabla n(r)|^2$ with constant $F_0$.
The EOS energy is 
\begin{equation}
W_{EOS} = \int d^3 r \ \epsilon_0 \left(n_n(r),n_p(r)\right),
\label{WEOS_def}
\end{equation}
and the gradient energy is
\begin{equation}
W_g = \int d^3 r \ F_0 |\nabla n(r)|^2.
\label{Wg_def}
\end{equation}
Note that the surface energy comes from both $W_{EOS}$ and $W_{g}$. See Appendix \ref{notes_on_oya1993} for the other choices of the inhomogeneity energy densities used in our early study of neutron star matter\cite{oya1993}.

The Coulomb energy is given by
\begin{equation}
W_C =\frac{e^2}{2} \int d^3 rd^3 r' \frac{n_p(r) n_p(r')}{|r-r'|}
\label{WC_def}
\end{equation}
with the electron charge $e$.

The OI model is a compressible liquid drop model capable of independently choosing radii and surface thicknesses for the neutron and proton distributions. We consider the point nucleon distribution $n_i(r) \ (i=n,p)$ as  a parametrized function of the distance $r$ from the center  with  edge radius parameter $R_i$ and relative surface diffuseness parameter $t_i$;
\begin{eqnarray}
n_i(r) = \begin{cases}
n_i^{in} \left(1 - \left(\frac{r}{R_i}\right)^{t_i}\right)^3 \qquad \qquad \qquad (r \le R_i) \\
0  \qquad \qquad \qquad \qquad \qquad \qquad (r \ge R_i), \end{cases}
\label{density_parametrization}
\end{eqnarray}
where $n_i^{in}$ is the central  density. The density at $r \ge R_i$ is zero because the nucleon density outside the classical turning point is zero. 

For given $Z$ and $A$, the values of the distribution parameters, $n_i^{in}$, $R_i$, and $t_i$, are chosen to minimize the mass excess calculated from Eqs. (\ref{Mex_def})--(\ref{WC_def}).
Equation (\ref{density_parametrization}) enables us to calculate the gradient energy $W_g$ and the Coulomb energy $W_C$ analytically.  

The root-mean-square (rms) radius of the charge distribution is given by
\begin{equation}
R_{ch}=\sqrt{\frac{1}{Z} \int d^3 r \ r^2 \int d^3 r' \ n_p(\bf{r'}) \rho(|\bf{r}-\bf{r}'|)},
\label{rms_charge_radius}
\end{equation}
using the proton charge form factor \cite{EltonSwift67};
\begin{equation}
\rho(r)=\left(\frac{1}{\sqrt{\pi} a_p}\right)^3 \exp \left[-(r/a_p)^2 \right],
\label{p_charge_form_factor}
\end{equation}
with $a_p=0.65$ (fm).  The rms radii of the proton and neutron distributions, $R_{rms\_p}(=R_{ch})$ and  $R_{rms\_n}$, are also calculated in the same way using the form factor $\rho(r)$. Then, the rms radius of the matter distribution, $R_m$, is given by
\begin{equation}
R_m=\sqrt{\frac{N \ R_{rms\_n}^2 + Z \ R_{rms\_p}^2}{A}} .
\end{equation}

The definition of surface thickness is not unique. The 90\%-10\% surface thickness is the distance between the surfaces where the density is 90\% and where it is 10\% of the central density. For the point nucleon distribution (\ref{density_parametrization}), the 90\%-10\% surface thickness is given by
\begin{equation}
                \textrm{thick}(i)=R_i \left[(1-0.1^ {1/3})^{1/t_i} - (1-0.9^ {1/3})^{1/t_i}\right] \quad (i=n,p).
\label{thickness90_10}
\end{equation}
We will use this quantity (\ref{thickness90_10}) to discuss the point nucleon distributions of the most stable nucleus in Sec. \ref{MSI_OI_ILD}.

\subsection{Optimization of interaction parameters}
\label{2017OIEOS_optimization}

The values of the five potential parameters $a_1-a_3$ and $b_1-b_2$, and the inhomogeneity parameter $F_0$ are optimized to fit the smoothed empirical data of neutron excess $I$, mass excess $M_{ex}$, and rms charge radius $R_{ch}$ of stable nuclei in Table \ref{empirical_data_tbl}.
In his early mass formula study \cite{MYamada1964}, Yamada evaluated the smoothed values of $I$ and $M_{ex}$ for the nine mass numbers based on the systematics of the neutron and proton separation energies \cite{YamadaMatumoto1961JPSJ} to represent the average trends of the $\beta$-stable nuclei. Note that the fractional $I$  values are allowed in Table \ref{empirical_data_tbl}. Meanwhile, the present author evaluated the $R_{ch}^{emp}$ values \cite{oya1993} from the rms charge radius data in Ref. \cite{1987ChargeRadii}. The empirical data in Table \ref{empirical_data_tbl} were used in our previous work \cite{OI2003} and 
our early neutron star matter study \cite{oya1993}. 
The optimization using the smoothed empirical data in Table \ref{empirical_data_tbl} gives an alternative way to obtain the EOS, presumably comparable to fitting the latest experimental mass data of all stable nuclei using a phenomenological nuclear interaction.
Numerically, the experimental data of $I$ and $M_{ex}$ for $\beta$-stable nuclei were known at the time of the evaluation in Ref. \cite{MYamada1964} with sufficient accuracy \cite{YamadaMatumoto1961JPSJ,1961NuclidicMassTable}. 
We also mention that deriving the macroscopic nuclear properties requires a certain smoothing or averaging, equivalent to evaluating the shell effects, which is more or less uncertain and dependent on phenomenological interactions.

%

The following empirical constraints are also imposed to limit the parameter space reasonably\cite{oya1993, OI2003}:
\begin{itemize}
\item many-body energy parameter of neutron matter ($b_3=1.58632 \ (\mathrm{fm}^3)$),
\item incompressibility of symmetric matter ($K_0=180,190, \cdots, 360 (\mathrm{MeV})$, 19 values),
\item slope of saturation curve at $\alpha=0$ ($-y=200,210, \cdots, 1800 \ (\mathrm{MeV \cdot fm^3})$, 16 values).
\end{itemize}

In  the present update, $-y=210, 230, 270  (\mathrm{MeV \cdot fm^3})$ are added to the previous version  \cite{OI2003} so that the present update has 304$(=16 \times 19)$  interactions while the earlier version has 247$(=13 \times 19)$  \cite{OI2003}.

\begin{table}[t]
\caption{The empirical values of the neutron excess $I^{emp}$, mass excess $M_{ex}^{emp}$ and rms charge radius $R_{ch}^{emp}$ as  functions of mass number $A$ on the smoothed $\beta$ stability line.\cite{oya1993, OI2003} }
\begin{center}
\begin{tabular}{cccc}
\hline
$A$  & $I^{emp}$ & $M_{ex}^{emp}$ (MeV) & $R_{ch}^{emp}$ (fm)\\
\hline
25	&   0.18    &  -13.10	& 3.029 \\
47	&   3.29    &  -46.17	& 3.567 \\
71	&   7.61    &  -72.38	& 3.997 \\
105	&  14.83   &  -89.69	& 4.487 \\
137	&  22.71   &  -84.89	& 4.874   \\
169	&  31.31   &  -61.18	& 5.206  \\
199	&  39.78   &  -23.12	& 5.466    \\
225	&  46.64   &   21.22	& --\\
245	&  52.22   &   61.21	& --\\
\hline
\end{tabular}
\end{center}
\label{empirical_data_tbl}
\end{table}%

For a given $(y, K_0)$,  the values of the four parameters $n_0, w_0, S_0$, and $F_0$ are chosen to fit the empirical data in Table \ref{empirical_data_tbl}. 
We can easily judge whether or not the numerical optimization result is physical from the saturation parameter values. Then, the interaction parameters, $a_1-a_3$ and $b_1-b_2$, are calculated from $y,  K_0, n_0, w_0$, and $S_0$. Eventually, we can calculate any saturation parameter from $a_1-a_3$ and $b_1-b_3$.

Specifically, we minimize
\begin{equation}
\chi^2 = \sum \Big[\Big(\frac{I^{cal} -I^{emp}}{\Delta I}\Big)^2 + \Big(\frac{M_{ex}^{cal} - M_{ex}^{emp}}{\Delta M}\Big)^2+   \Big(\frac{R_{ch}^{cal} - R_{ch}^{emp}}{\Delta R}\Big)^2 \Big] , 
\label{chi2}
\end{equation}
with $\Delta I =0.1, \ \Delta M =1$ (MeV) and $\Delta R=0.01$ (fm).  This optimization is not easy because we first calculate (optimize)  the most stable isobars for the nine mass numbers in Table \ref{empirical_data_tbl} and then optimize the $\chi^2$ value in Eq.  (\ref{chi2}).  In the present update, the initial values of the parameters are chosen to make the optimum parameter values vary smoothly as functions of $(y, K_0)$.

\begin{figure}[htbp]
\begin{center}
\includegraphics[width=15cm]{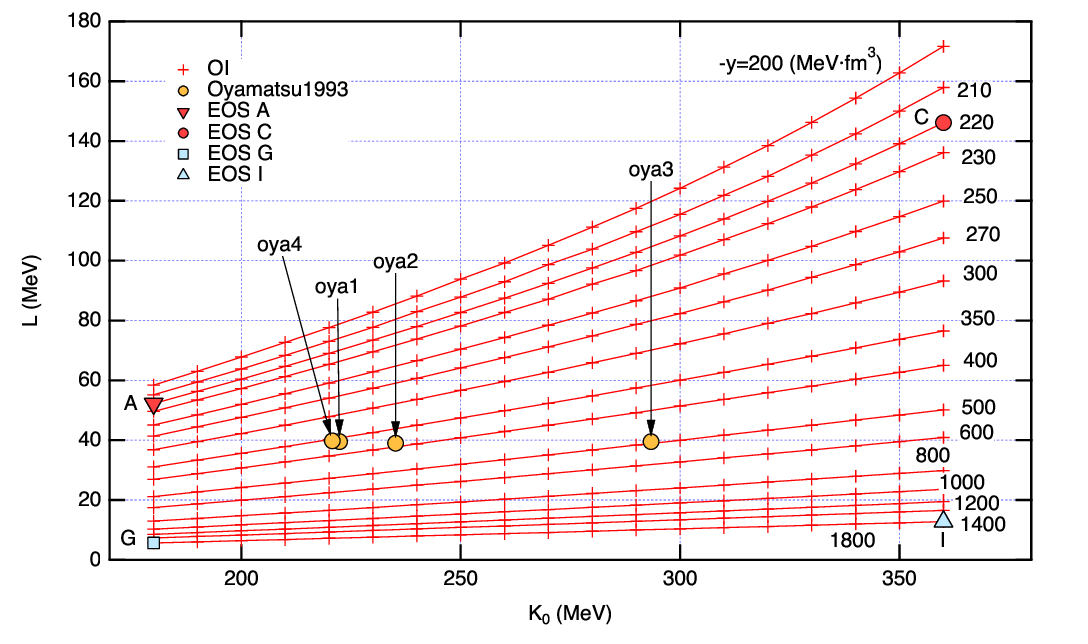}
\caption{The plots of  $(K_0, L)$  for the present 304 interactions. The value of the slope $y$ is attached to the line joining the points with the same $y$ value. Also plotted are the four models oya1-4 in our early neutron star matter study \cite{oya1993} and the four extreme EOSs A, C, G,  and I defined in our previous study \cite{OI2003} (see also Table \ref{EOS_ACGI_oya93}). }
\label{y_K0L_170313}
\end{center}
\end{figure}

\begin{table}[t]
\caption{The values of saturation parameters of four models I-IV (oya1-4) in our early study of neutron-star matter \cite{oya1993}, together with the present update values of four extreme EOSs (A, C, G, and I) defined in Ref. \cite{OI2003}. The $S_0$ values are calculated using Eq. (\ref{S_n_def}). Note that the inhomogeneity energy of  oya1-3 and the  definition of $S_0$ were different in the early study\cite{oya1993} (see Appendix \ref{notes_on_oya1993}).}
\begin{center}
\begin{tabular}{cccccccc}
\hline
EOS & $-y$ & $K_0$  &  $L$ & $n_0$ & $w_0$ & $S_0$ & $F_0$ \\
  & (MeV $\cdot$ fm$^3$)   &  (MeV)    & (MeV)  & (fm$^{-3}$)  &  (MeV)   & (MeV)  & (MeV $\cdot$  fm$^5$) \\
\hline
oya1 & 359.93 & 222.41  & 39.559 & 0.15856 & -16.076 & 30.452 & 47.399 \\
oya2 & 411.63 & 235.14  & 39.010 & 0.15227 & -16.013 & 31.195 & 49.522 \\
oya3 & 479.15 & 293.36  & 39.513 & 0.15845 & -16.312 & 30.678 & 47.294 \\
oya4 & 359.27 & 220.76  & 39.743 & 0.15807 & -16.070 & 30.671 & 68.650 \\
A & 220 & 180  & 52.266 & 0.16921 & -16.252 & 32.427 & 71.360 \\
C & 220 & 360  & 146.16 & 0.14578 & -16.119 & 39.065 & 66.985 \\
G & 1800 & 180  & 5.6552 & 0.16864 & -16.189 & 28.611 & 69.856 \\
I & 1800 & 360  & 12.789 & 0.14896 & -16.031 & 28.575 & 61.660 \\
\hline
\end{tabular}
\end{center}
\label{EOS_ACGI_oya93}
\end{table}%

Figure \ref{y_K0L_170313} plots the $(K_0, L)$ values for the 304 interactions (EOSs) and shows lines joining the points with the same $y$ value for the eye guide. The $L$ value is calculated from Eq. (\ref{slope_y}) using $y, n_0, S_0$, and $K_0$. 
This figure shows the one-to-one correspondence between $(K_0, y)$ and $(K_0, L)$. Hereafter, we take $L$ as an independent parameter instead of $y$ and analyze the interaction and saturation parameters as functions of $(K_0, L)$. Figure \ref{y_K0L_170313} also depicts the four models oya1-4 of our early neutron-star matter study\cite{oya1993} and four extreme EOSs A, C, G, and I defined in our previous work \cite{OI2003}, whose values of saturation parameters are listed in Table  \ref{EOS_ACGI_oya93}. The neutron matter EOSs of oya1-4 have only one free potential parameter because the study fixed the $b_1/b_2$ value. This constraint leads to $L \approx 40$ MeV, close to the FP EOS fit \cite{oya1993}.

\begin{figure}[tb]
\begin{center}
\includegraphics[width=12cm]{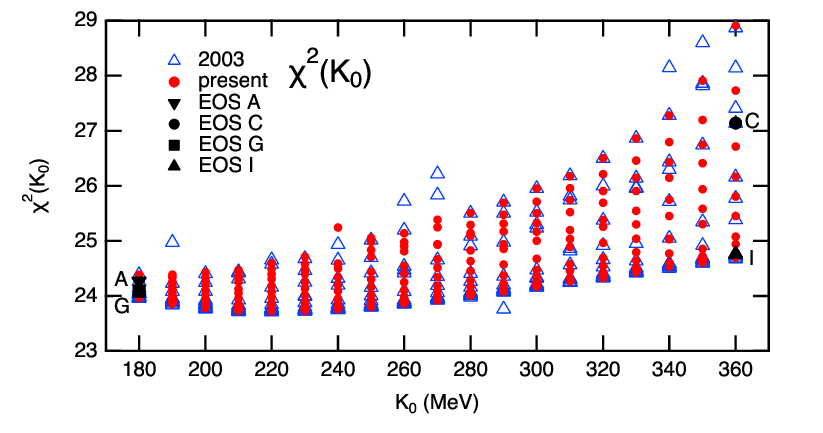}
\includegraphics[width=12cm]{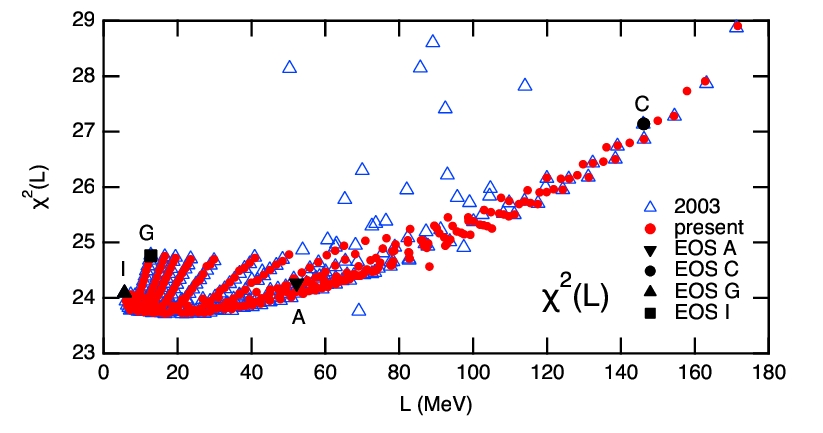}
\caption{The $\chi^2$ values of the present update (filled red circles) and the previous 2003 study (open triangles) \cite{OI2003} as  functions of $K_0$ and $L$. Also plotted are EOSs A, C, G, and I \cite{OI2003} in Table \ref{EOS_ACGI_oya93}.}
\label{fig_chi_sq}
\end{center}
\end{figure}

Figure \ref{fig_chi_sq} plots the optimum $\chi^2$ values for the present  304 interactions  and the previous 247 interactions \cite{OI2003}. 
The range of  $\chi^2$  is relatively narrow, and  $\chi^2$ is minimum at $K_0=220$ MeV and $L$=16.580 MeV.
The present optimization reasonably minimizes the $\chi^2$ value and improves the overall optimization results compared with the previous work  \cite{OI2003}. Meanwhile, the previous version shows insufficient minimizations and overfitting at $K_0=290$ (MeV) and $L\approx69$ (MeV).  These kinds of numerical uncertainty are inevitable even in the present update. To overcome these difficulties, we take many data points and focus on gross behavior as a function of $(K_0, L)$ in our studies \cite{OI2003,  OI2007, Oyamatsu:2010bf, Oyamatsu:2010sk, Iida:2013fra, Sotani:2012qc, Sotani:2012xd, Sotani:2013jya, Sotani:2013dga, Sotani:2015lya, Sotani:2015laa, Sotani:2016pmb, Sotani:2017hpq, Sotani:2018tdr, Sotani:2019pja}.

\begin{figure}[htbp]
\begin{center}
\includegraphics[width=12cm]{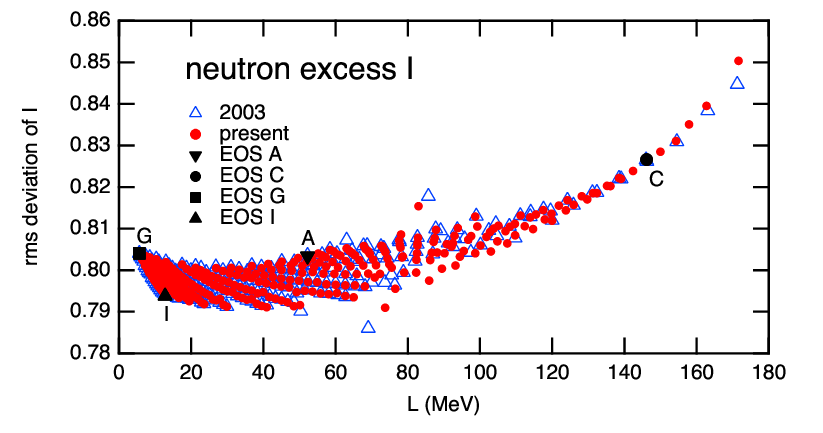}
\includegraphics[width=12cm]{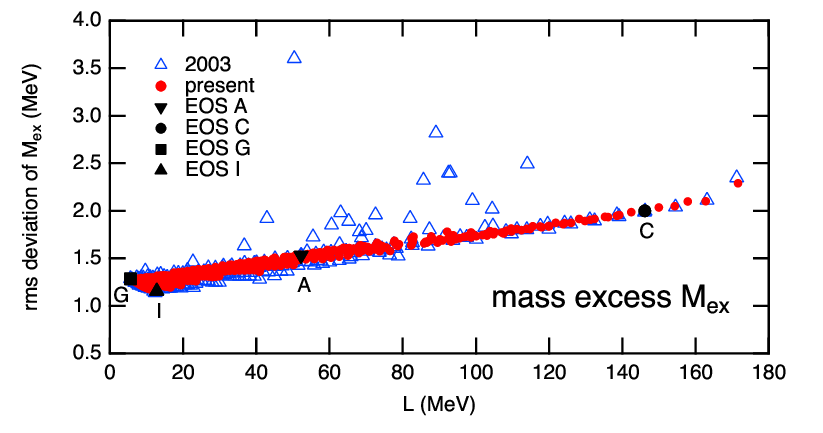}
\includegraphics[width=12cm]{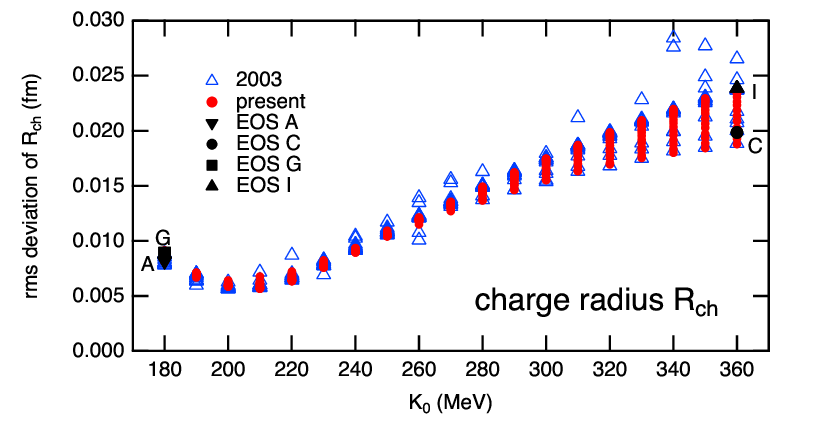}
\caption{The rms deviations of neutron excess $I$, mass excess $M_{ex}$, and charge radius $R_{ch}$ from the 
smoothed 
empirical values in Table \ref{empirical_data_tbl}. The symbols are the same as those in Fig. \ref{fig_chi_sq}.}
\label{fig_rms_dev_IMR}
\end{center}
\end{figure}

To examine the significant contribution to the optimum $\chi^2$, Fig. \ref{fig_rms_dev_IMR} shows the rms deviations of neutron excess $I$, mass excess $M_{ex}$, and charge radius $R_{ch}$ from the empirical values in Table \ref{empirical_data_tbl}. Their correlations with $K_0$ and $L$ are seen more clearly in the present study. From Eq. (\ref{chi2}) and Fig.  \ref{fig_rms_dev_IMR}, we see that the dominant contribution to $\chi^2$ comes from neutron excess $I$. The rms deviation of neutron excess $I$ mainly correlates with $L$ and has appreciable sensitivity with $K_0$, while that of mass excess $M_{ex}$ strongly correlates with $L$. Meanwhile, the rms deviation of charge radius $R_{ch}$ strongly correlates with $K_0$ below 280 (MeV); above this $K_0$ value, its sensitivity to $L$ increases with $K_0$. It is also noted that the overfitting of $\chi^2$ in Fig. \ref{fig_chi_sq} at $K_0=290$ (MeV) and $L\approx69$ (MeV) stems from the overfitting of neutron excess $I$ in Fig. \ref{fig_rms_dev_IMR}. 

It is remarked that the present 304 interactions fit the empirical values of $I^{emp}$, $M_{ex}^{emp}$, and $R_{ch}^{emp}$  almost equally in the sense that the deviations from the empirical values are much smaller than the fluctuations due to the shell effects, as shown in Fig. \ref{fit_I_Mex_Rch} in Appendix \ref{comp_cal_exp}. Thus, the $K_0$ and $L$ degrees of freedom are nearly frozen for $I$, $M_{ex}$, and $R_{ch}$  of stable nuclei. We will confirm this insensitivity for $M_{ex}$ 
 in Sec. \ref{nuclear_masses_and_EOS}.

\clearpage

\section{Numerical results}
\label{OIEOSF2017_interaction_saturation_paramter_values}

\subsection{Optimum values of potential parameters and their correlations}
\label{correlations_interaction_parameters}
\begin{figure}[htbp]
\begin{center}
\includegraphics[width=9.5cm]{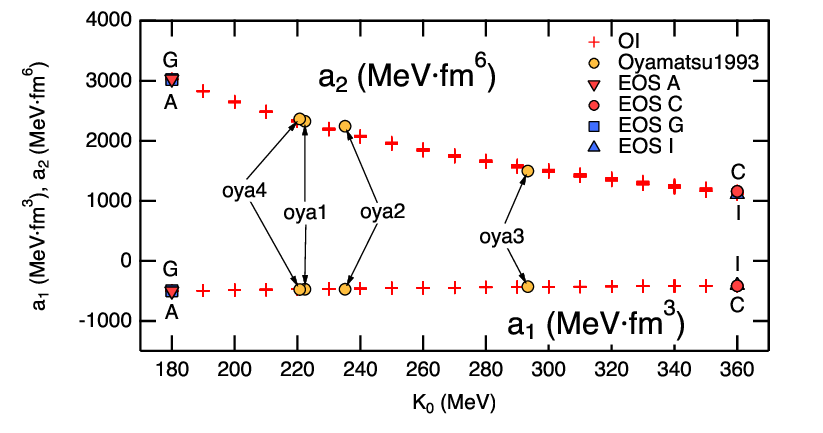}
\includegraphics[width=9.5cm]{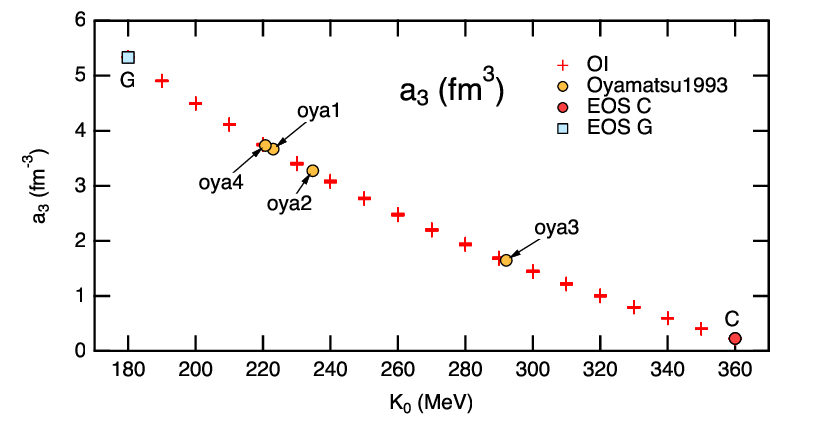}
\includegraphics[width=9.5cm]{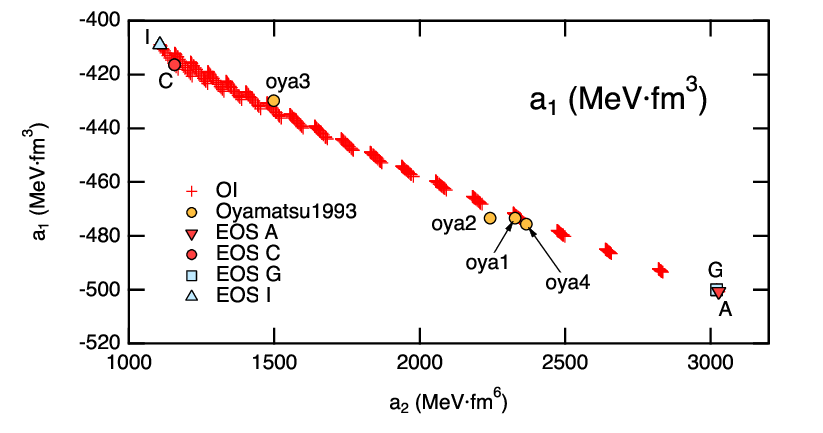}
\includegraphics[width=9.5cm]{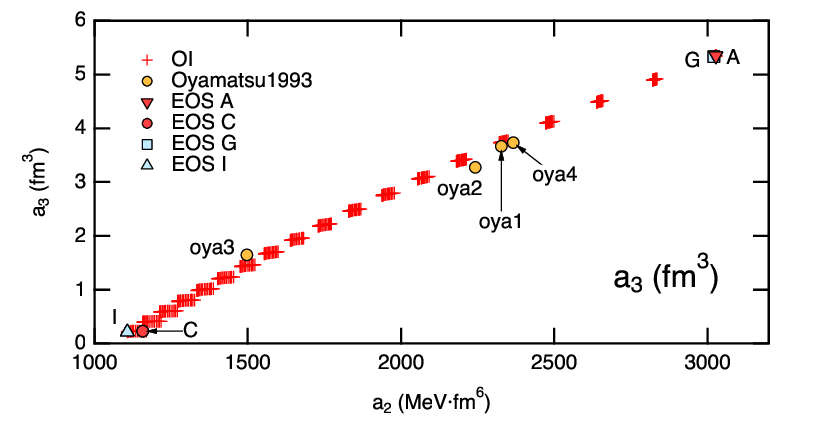}
\caption{The potential parameters $a_1 - a_3$ as functions of $K_0$ (upper),  and $a_1, a_3$ as functions of $a_2$ (lower). Also plotted are the four models (oya1-4) in the early study \cite{oya1993}, and EOSs A, C, G, and I in Table \ref{EOS_ACGI_oya93}.}
\label{fig_abK0L}
\end{center}
\end{figure}

The symmetric matter potential, $v_s(n)$, essentially has only one degree of freedom.
Figure \ref{fig_abK0L}  shows, in the upper panels, strong correlations among the potential parameters  $a_1 - a_3$ and the incompressibility $K_0$. The potential parameters $a_1 - a_3$ show clear dependences on $K_0$. The lower panels show that the two-body energy coefficient $a_1$ and the many-body parameter $a_3$ strongly correlate with the three-body energy coefficient $a_2$.  Consequently, the symmetric matter EOS $w_s(n)$ depends on $K_0$ and the three-body energy coefficient $a_2$.

\begin{figure}[htbp]
\begin{center}
\includegraphics[]{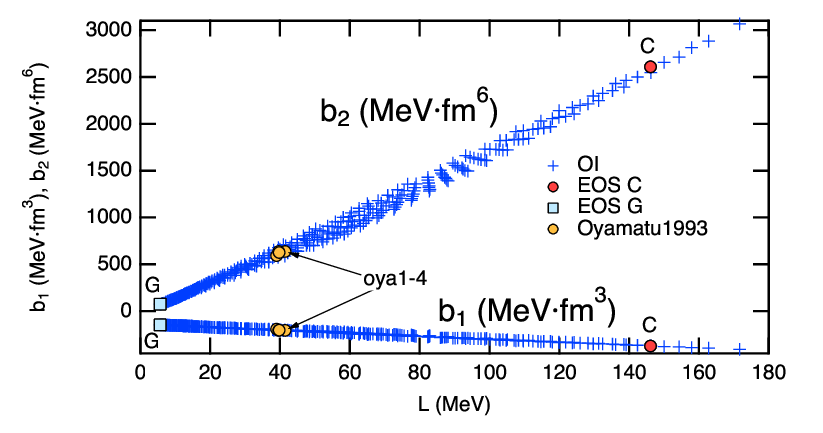}
\includegraphics[]{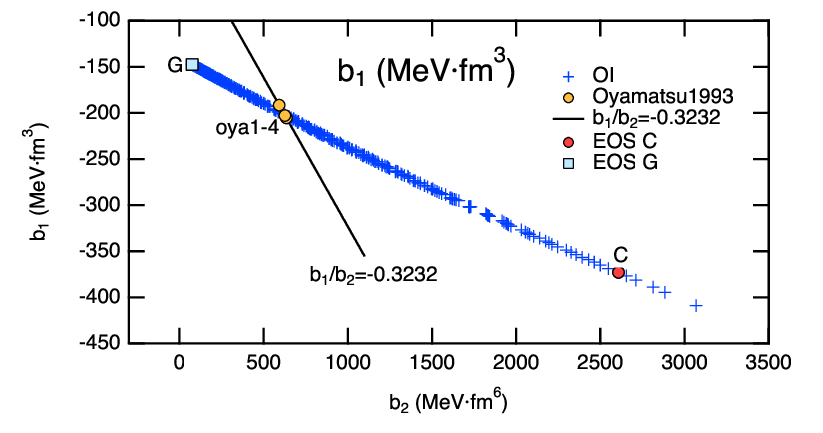}
\caption{The potential parameters $b_1$ and $b_2$ as functions $L$ (upper), and $b_1$ as function of $b_2$ (lower). Also plotted are the four models (oya1-4) in the early study \cite{oya1993} and EOSs C and G in Table \ref{EOS_ACGI_oya93}.}
\label{fig_ab}
\end{center}
\end{figure}

Similarly, the neutron matter potential, $v_n(n)$, essentially has only one degree of freedom.
Figure \ref{fig_ab} shows that, in the upper panel, the potential parameters $b_1$ and $b_2$ have clear dependences on $L$ while, in the lower panel, the two-body energy coefficient $b_1$ also correlates strongly with the three-body energy coefficient $b_2$.  Consequently, the neutron matter EOS $v_n(n)$ depends on $L$ and the three-body energy coefficient $b_2$.

%

It is also noteworthy in Figs. \ref{fig_abK0L} and \ref{fig_ab} that the two-body coefficients ($a_1, b_1$) are  constrained much better than the three-body coefficients ($a_2, b_2$) and the many-body coefficient $a_3$. Meanwhile, the $b_3$ value, relevant to the high-density EOS, is so uncertain that we fix the value in our studies \cite{oya1993,OI2003,OI2007,Oyamatsu:2010bf,Oyamatsu:2010sk,Iida:2013fra,Sotani:2013dga,Sotani:2012qc,Sotani:2012xd,Sotani:2013jya, Sotani:2015lya, Sotani:2015laa, Sotani:2016pmb,Sotani:2017hpq,Sotani:2018tdr, Sotani:2019pja}. 

\clearpage

\subsection{Values of saturation parameters and their correlations}
\label{correlations_saturation_parameters}
\begin{figure}[htbp]
\begin{center}
\includegraphics[width=12cm]{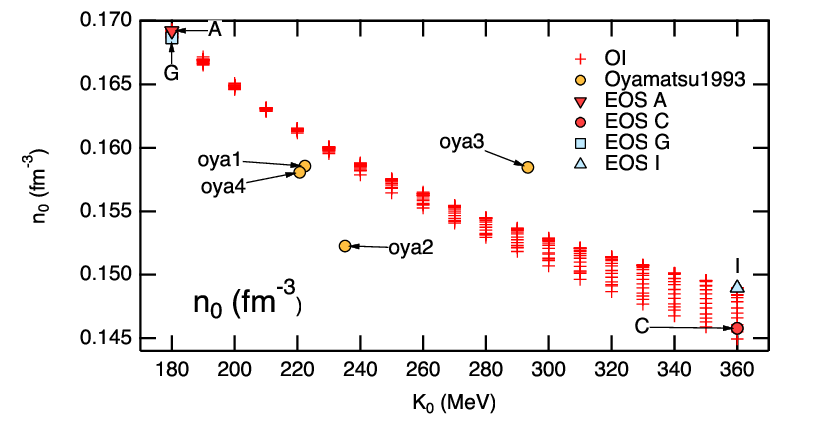}
\includegraphics[width=12cm]{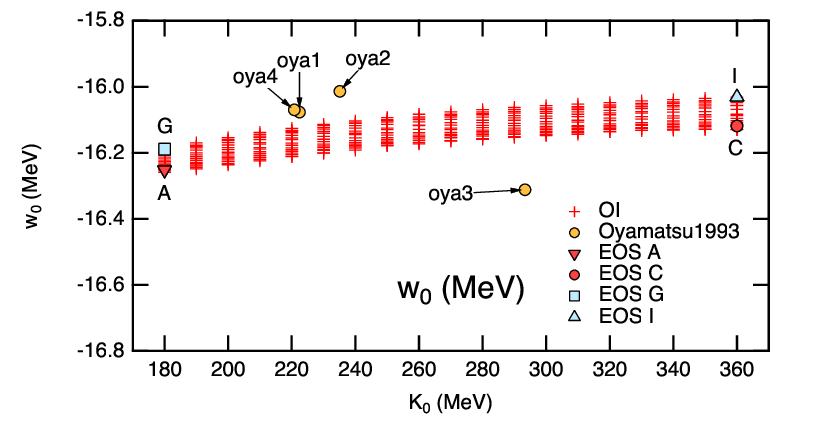}
\includegraphics[width=12cm]{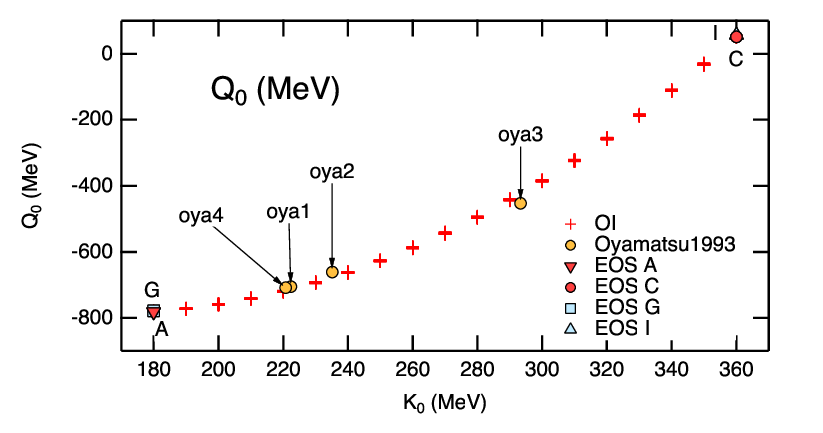}
\caption{The $K_0$ correlations of the saturation parameters $n_0, w_0$ and $Q_{0}$. Also plotted are the four models (oya1-4) in the early study \cite{oya1993}, and EOSs A, C, G, and I in Table \ref{EOS_ACGI_oya93}.}
\label{fig_n0w0Q0_correlations}
\end{center}
\end{figure}

Figure \ref{fig_n0w0Q0_correlations} shows the $K_0$ correlations of the saturation parameters $n_0, w_0$, and $Q_{0}$ of symmetric matter.
Naturally, these parameters correlate with $K_0$ because the potential parameters $a_1-a_3$ of the symmetric matter EOS $w_s(n)$ correlate with  $K_0$.
The saturation density  $n_0$ shows a relatively clear correlation  with $K_0$ in Fig. \ref{fig_n0w0Q0_correlations}, except for appreciable sensitivities to $L$ at $K_0 \gtrsim$300 MeV.
On the other hand, the saturation energy $w_0$ is well constrained within about $\pm 0.1$ MeV and has subtle but relatively clear sensitivity to  $K_0$ and $L$.   This sensitivity is not negligible in the sense that 0.05 MeV/nucleon difference in $^{208}$Pb amounts to a 10 MeV difference of its mass excess.

For $Q_0$, we also see its clear correlation with $K_0$ in Fig. \ref{fig_n0w0Q0_correlations}. Note  that neither  $K_0$ nor $Q_0$ includes the two-body energy coefficient $a_1$, so  we expect  a simple relation between $K_0$ and $Q_0$ in  the OI model  (see  Eqs.  (\ref{K0_def}) and (\ref{Q0_def})). Eventually, except for the subtle $L$ dependences for $n_0$ ($K_0 \gtrsim 300$ MeV) and $w_0$, we confirm again that the symmetric matter EOS mainly depends on $K_0$.

\begin{figure}[htbp]
\begin{center}
\includegraphics[width=9.5cm]{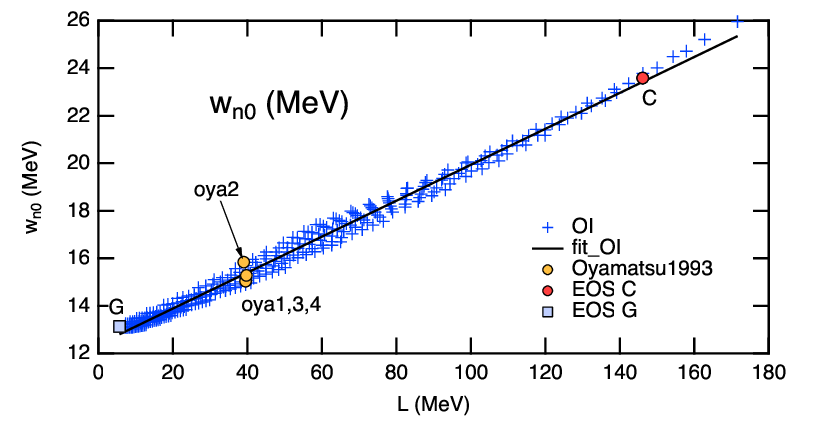}
\includegraphics[width=9.5cm]{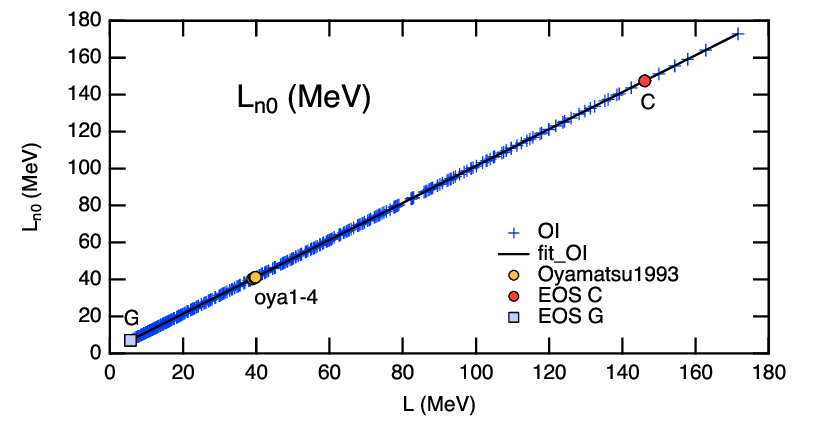}
\includegraphics[width=9.5cm]{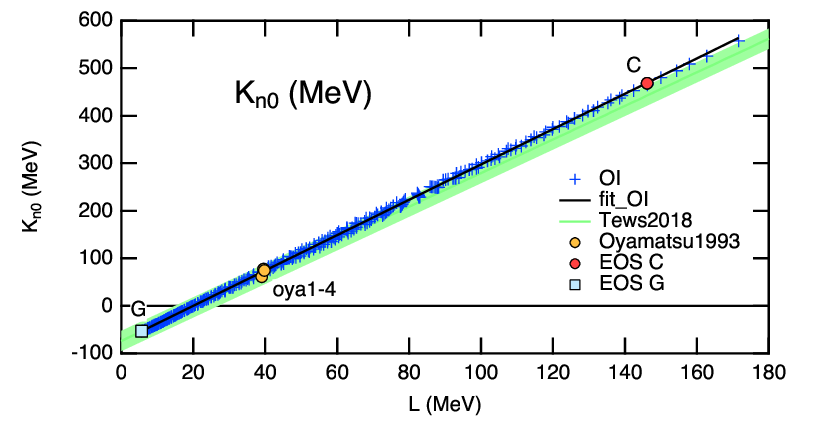}
\includegraphics[width=9.5cm]{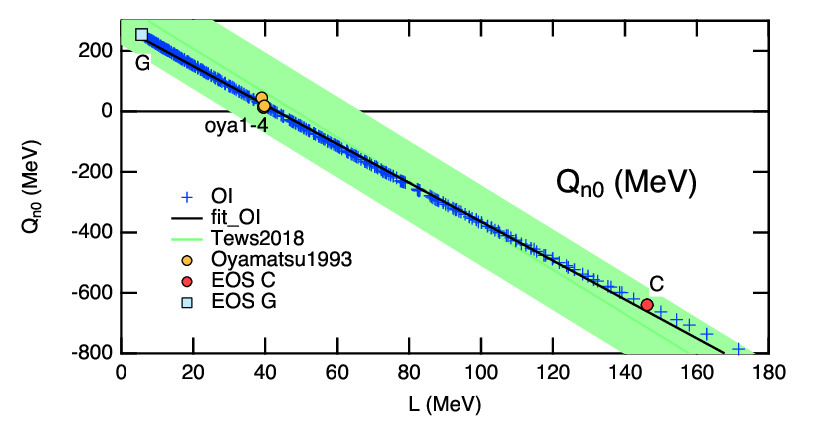}
\caption{The $L$ correlations of the saturation parameters $w_{n0}, L_{n0}, K_{n0}$, and $Q_{n0}$. Their fitting lines (fit\_OI) are also shown. The shaded areas (Tews2018) for $K_{n0}$ and $Q_{n0}$  are calculated with the fitting formulae by Tews et al. \cite{TewsLattimerOhnishiKolomeitsev2018}, which enclose 68.3\% of their accepted interactions. We also plot the four models (oya1-4) in the early study \cite{oya1993} and EOSs C and G in Table \ref{EOS_ACGI_oya93}. }
\label{fig_wn0Ln0Kn0Qn0_correlations}
\end{center}
\end{figure}

Figure \ref{fig_wn0Ln0Kn0Qn0_correlations} shows the $L$ correlations of saturation parameters $w_{n0}, L_{n0}, K_{n0}$, and $Q_{n0}$ of neutron matter. 
Naturally, these  parameters correlate with $L$ because the potential parameters $b_1-b_3$ of the neutron matter EOS $w_n(n)$ correlate with $L$. We see their strong correlations with $L$ in Fig. \ref{fig_wn0Ln0Kn0Qn0_correlations} and obtain the following fitting formulae:
\begin{equation}
w_{n0} = 12.367 \pm  0.0264 + (0.075639 \pm  0.000404) L \  \textrm{(MeV)},
\label{wn0-L_correlation}
\end{equation}
\begin{equation}
L_{n0} = 1.3611 \pm  0.00316 + (0.99956 \pm 4.85 \times 10^{-05}) L  \ \textrm{(MeV)},
\label{Ln0-L_correlation}
\end{equation}
\begin{equation}
K_{n0} = -74.636 \pm 0.281 + (3.7193 \pm  0.00431) L \  \textrm{(MeV)},
\label{Kn0-L_correlation}
\end{equation}
\begin{equation}
Q_{n0}  = (278.84 \pm 0.739) + (-6.4345 \pm 0.0113) L \  \textrm{(MeV)}.
\label{Qn0_L_correlation_OI}
\end{equation}
The difference between $L_{n0}$ and $L$ is only 1 MeV, which is the kinetic energy of higher orders than $\alpha^2$. Finally, we remark that we have a simple relation between  $K_{n0}$ and $Q_{n0}$ because the three-body coefficient $b_2$ is only one free parameter in Eqs. (\ref{Kn0_def}) and (\ref{Qn0_def}) (the $b_3$ value is fixed).

\begin{figure}[htbp]
\begin{center}
\includegraphics[width=12cm]{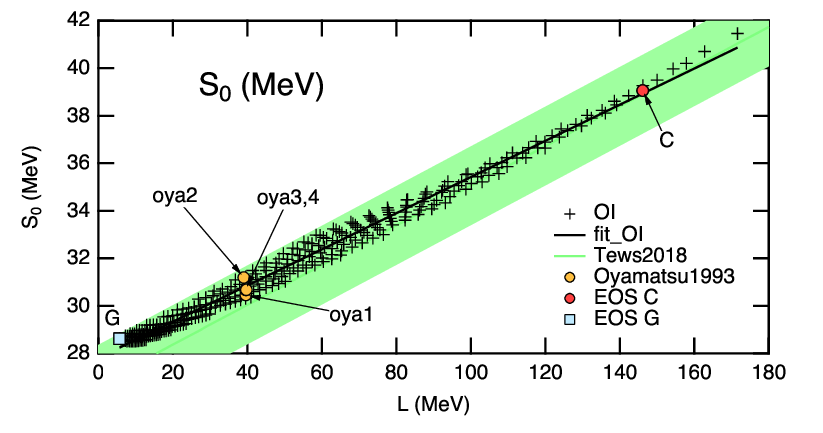}
\includegraphics[width=12cm]{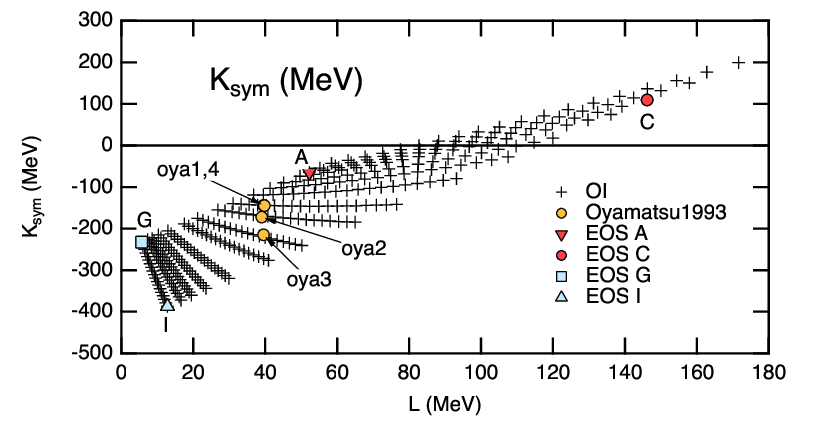}
\includegraphics[width=12cm]{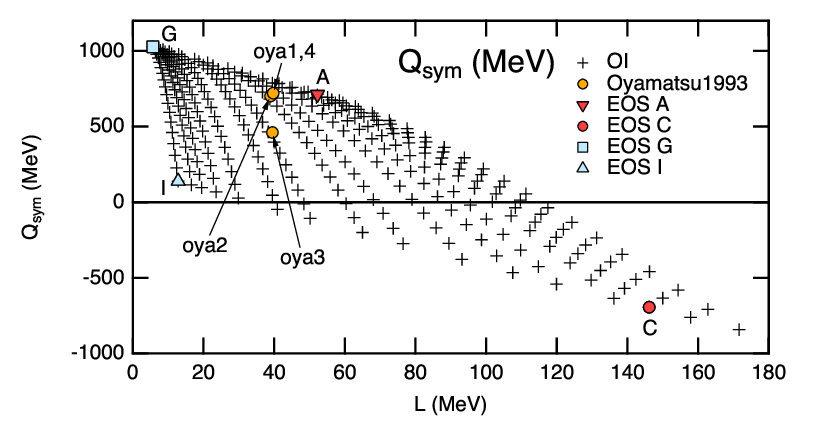}
\caption{The $L$ correlations of the saturation parameters $S_0, K_{sym}$, and $Q_{sym}$. Also plotted are the four models (oya1-4) in the early study \cite{oya1993} and EOSs A, C, G, and I in Table \ref{EOS_ACGI_oya93}. The shaded area for  $S_0$ is calculated with the fitting formula by Tews et al. \cite{TewsLattimerOhnishiKolomeitsev2018}, which encloses 68.3\% of their accepted interactions.}
\label{fig_S0KsymQsym_correlations}
\end{center}
\end{figure}

Figure \ref{fig_S0KsymQsym_correlations} shows the $L$ correlations of the saturation parameters $S_0, K_{sym}$, and $Q_{sym}$, of the density-dependent symmetry energy $S(n)$. 
Except for $S_0$, the saturation parameters of $S(n)$ also have clear $K_0$ dependence from $w_s(n)$ because $S(n) \approx w_n(n)-w_s(n)$.
The symmetry energy $S_0$ has a simple correlation with $L$ and is  approximately given by
\begin{equation}
S_0 = 27.809 \pm 0.0291 + (0.0761 \pm  0.000446) L  \  \textrm{(MeV)}.
\label{S0-L_correlation}
\end{equation}
The values of the coefficients in Eq. (\ref{S0-L_correlation}) are slightly different but essentially the same as those in our previous study \cite{OI2003}. 
This simple correlation (\ref{S0-L_correlation}) stems from the fact that the saturation energy $w_0$ is essentially constant of $K_0$ and $L$ so that $S_0 \approx w_{n0}-w_0$ reflects the $L$ dependence in $w_{n0}$ except for subtle $K_0$  dependence.

The saturation parameters  $K_{sym}$ in Eq. (\ref{Ksym_def}) and $Q_{sym}$  in Eq. (\ref{Qsym_def})  have relatively complicated dependence on $K_0$ and $L$  because they include both $K_0$-dependent potential parameters $a_2-a_3$ and $L$-dependent $b_2$ (the value of $b_3$ is fixed). It is also noted that $K_{sym}$ and $Q_{sym}$ do not include two-body parameters.

The shaded areas for $K_{n0}, Q_{n0}$, and $S_0$ in Figs. \ref{fig_wn0Ln0Kn0Qn0_correlations} and  \ref{fig_S0KsymQsym_correlations}  are  calculated with fitting formulae by Tews et al. \cite{TewsLattimerOhnishiKolomeitsev2018}, which enclose 68.3\% of their accepted 188 Skyrme and 73 RMF interactions. These figures show that the saturation parameter values of neutron matter EOS in the OI model are consistent with the general trends of the phenomenological interactions.

\subsection{Fixed points of $S(n)$ and $w_n(n)$}
The $S_0 - L$ correlation in Eq. (\ref{S0-L_correlation}) implies that the symmetry energy should have a reasonable value at the nuclear surface.
Actually, in the lowest approximation,
\begin{equation}
S(n) \approx  S_0 + u L = 27.809+(0.0761+u) L.
\label{Sn_S0-L}
\end{equation}
The symmetry energy at $u=-0.0761$ ($n/n_0 = 0.7717$)  is constant (27.809 MeV) independently of $L$.
Thus with the $S_0 - L$ correlation (\ref{S0-L_correlation}), we have practically only one degree of freedom for $S(n)$ at $n \simeq n_0$ and choose the slope $L$ as the independent EOS parameter to study the nuclear structure. 

Similarly, from the $w_{n0}-L$ correlation (\ref{wn0-L_correlation}) and the $L_{n0}-L$ correlation (\ref{Ln0-L_correlation}) for neutron matter, we have, in the lowest approximation,
\begin{equation}
w_n(n) \approx  w_{n0} + u L_{n0} \approx 12.367+ 1.3611u+ (0.075639+0.99956 u) L.
\end{equation}
Then, the neutron-matter energy at $u=-0.0757$ ($n/n_0 = 0.7730$) is constant (12.264 MeV) independently of $L$. This property is an empirical constraint of neutron matter EOS, which Brown discussed using phenomenological interactions \cite{ABrown2000}. We mention that this constraint is obtained only from  the empirical mass and radius data of stable nuclei in the present study.

\clearpage

\subsection{Inhomogeneity energy and saturation parameters}
\begin{figure}[htbp]
\begin{center}
\includegraphics[]{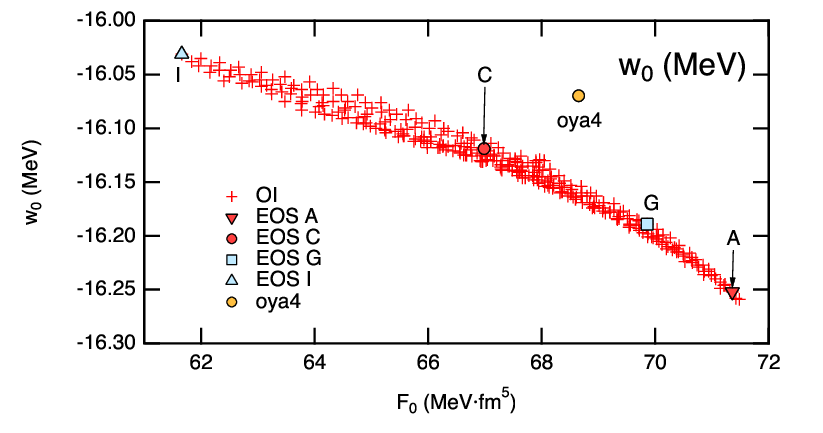}
\includegraphics[]{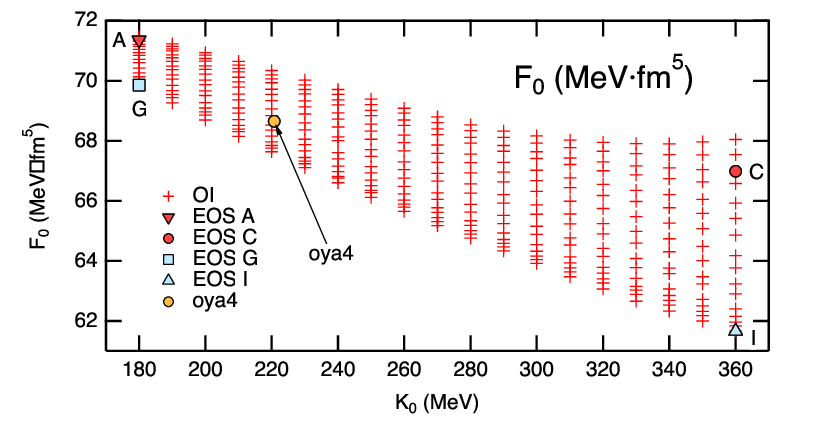}
\caption{The correlations between $w_0$ and $F_0$ (upper) and between $F_0$ and $K_0$ (lower). Also plotted are the oya4 in the early study \cite{oya1993} and EOSs A, C, G, and I  in Table \ref{EOS_ACGI_oya93}. }
\label{fig_w0F0_K0L}
\end{center}
\end{figure}

The $w_0-F_0$ relation in Fig. \ref{fig_w0F0_K0L} (upper) represents the correlation between uniform-matter and inhomogeneity energies of isoscalar interaction. Interestingly, the inhomogeneity energy parameter $F_0$ has complicated sensitivities to $K_0$ and $L$ (Fig. \ref{fig_w0F0_K0L} (lower)).
In contrast,  the saturation energy $w_0$ is almost constant, showing only subtle sensitivities to $K_0$ and $L$ in  Fig. \ref{fig_n0w0Q0_correlations}. Despite these different sensitivities to $K_0$ and $L$, 
the $w_0-F_0$ correlation is natural because the potential contribution of the inhomogeneity energy is well represented by gradient expansion, whose coefficients are the spacial moments of the long-range part of inter-nucleon potentials \cite{Brueckner:1968zzb}. In the OI model, we assume that the kinetic contribution is effectively included in the parameter $F_0$.

In principle, the choice of different inhomogeneity energy terms can make differences in the saturation density $n_0$ and energy $w_0$ because the inhomogeneity energy affects the local pressure equilibrium in a nucleus. Here we discuss the sensitivity of the inhomogeneity energy using the oya1-4 models of the early study \cite{oya1993} in Table \ref{EOS_ACGI_oya93} (see also  Appendix \ref{notes_on_oya1993}), keeping in mind that the optimization was probably poorer than the present one.
\begin{itemize}
\item The oya4 model has the same inhomogeneity energy as the OI model.
\item The oya1-3 models have the kinetic contribution of the inhomogeneity energy in addition to the potential contribution.
\item The oya2 model also includes an extremely large isovector gradient term.
\item The $K_0$ and $L$ ($-y$) values of the oya1-4 models were also optimized with the additional constraint $b_1/b_2=-0.3232$ (fm$^{-3}$).
\end{itemize}

The values of the potential and saturation parameters of all oya1-4 models in Figs. \ref{fig_abK0L}-\ref{fig_S0KsymQsym_correlations} agree well with those of the OI model except for the slight differences in the $n_0$ and $w_0$ values in Figs. \ref{fig_n0w0Q0_correlations} and \ref{fig_w0F0_K0L}.
Notably, the excellent agreement of the oya1 and oya4 results encourages our assumption that the kinetic contribution of the inhomogeneity can be effectively included in the coefficient $F_0$.
Furthermore, the neutron matter EOS parameters of all oya1-4 in Figs. \ref{fig_ab} and \ref{fig_wn0Ln0Kn0Qn0_correlations} agree very well ($L \approx 40$ MeV), so the neutron matter EOS seems insensitive to the inhomogeneity energy even with the large isovector inhomogeneity energy in the oya2 model.
Interestingly, the oya2 values of the symmetric matter EOS parameters in Figs. \ref{fig_abK0L} and \ref{fig_n0w0Q0_correlations} differ from the oya1 and oya4 values.
Hence, the isovector inhomogeneity energy cancels the isoscalar inhomogeneity energy and affects the symmetric matter EOS (see  Table \ref{EOS_ACGI_oya93} and Appendix \ref{notes_on_oya1993}). 



\subsection{Nuclear masses and the empirical EOSs}
\label{nuclear_masses_and_EOS}
\begin{figure}[htbp]
\begin{center}
\includegraphics[]{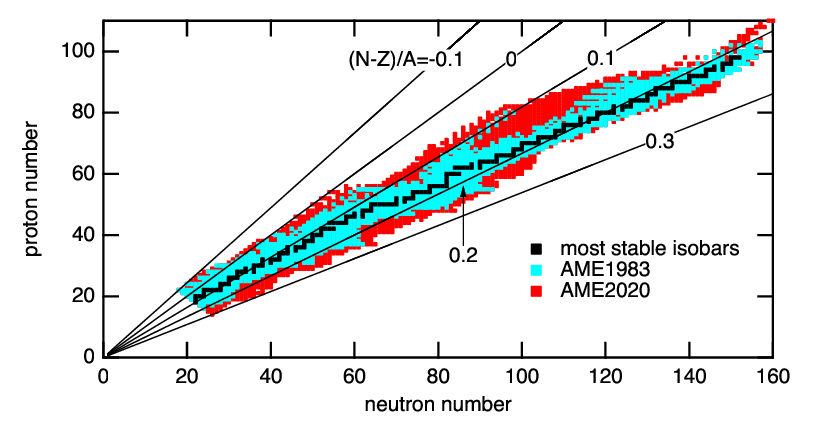}
\caption{The nuclides with $A\ge40$, whose masses are calculated and compared with the experimental values in AME1983\cite{Wapstra:1985hmr} and AME2020\cite{Wang:2021xhn}. We only calculate the masses of the most stable isobars with $40 \le A \le 250$.}
\label{fig_AME2020_nuclides}
\end{center}
\end{figure}

\begin{figure}[htbp]
\begin{center}
\includegraphics[]{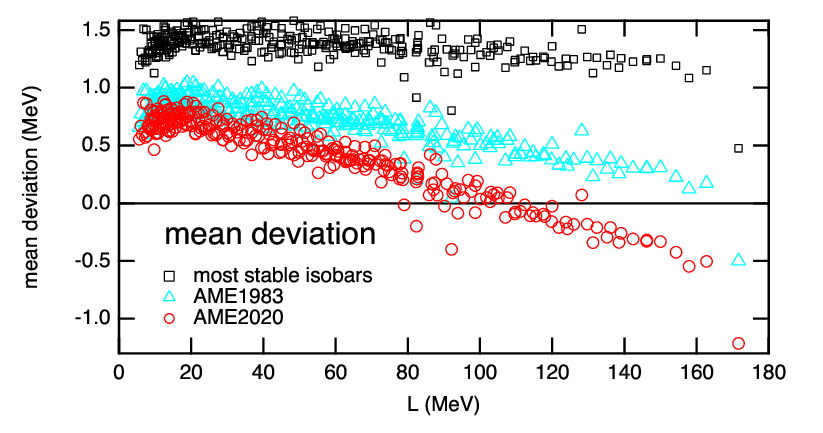}
\includegraphics[]{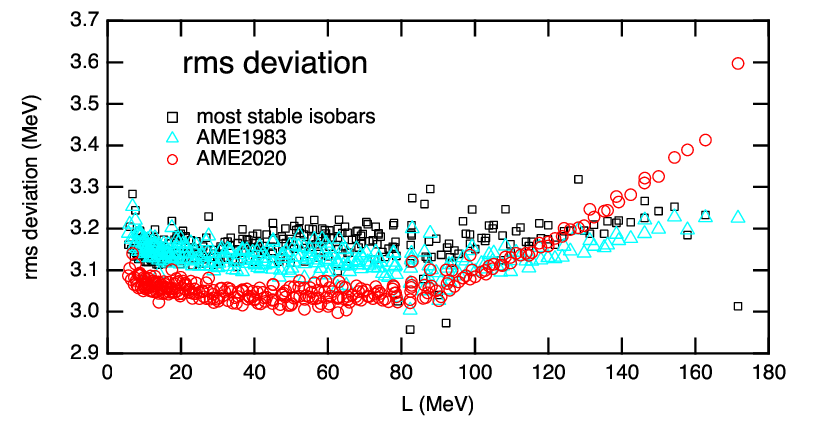}
\caption{The mean deviations (upper) and the rms deviations (lower) of the calculated masses from the experimental ones as functions of $L$. The nuclear mass calculations were performed for the most stable isobars, the AME1983 nuclides\cite{Wapstra:1985hmr} and  the AME2020 nuclides\cite{Wang:2021xhn} in Fig. \ref{fig_AME2020_nuclides}. }
\label{fig_AME2020_mass}
\end{center}
\end{figure}

Figure \ref{fig_AME2020_nuclides} shows the nuclides with $A \ge 40$, whose experimental mass values are compared with the calculated values using the OI model's update 304 interactions (EOSs);
2301 nuclides in Atomic Mass Evaluation 2020 (AME2020)\cite{Wang:2021xhn}, 1514 nuclides in AME1983\cite{Wapstra:1985hmr} together with 211 most stable isobars with $40 \leq A \leq 250$. 
We exclude the lighter nuclei because the OI model overestimates their masses, as shown in Fig. \ref{fig_mex_alpha_MSI} (upper).
Figure \ref{fig_AME2020_mass} shows the mean deviations (upper) and root-mean-square deviations (lower) from the experimental mass values in  AME2020 and AME1983.
The mean and rms deviations for the most stable isobars (MSIs) are also plotted in Figure \ref{fig_AME2020_mass}. 
The mean deviation for MSIs is almost constant of $L$ at about 1.2--1.6 MeV but shows appreciable scattering, presumably reflecting the strong shell effects in the MSIs. The rms deviation for the MSIs is also almost constant at about 3.1 -- 3.2 MeV and much larger than 1.1 -- 2.3 MeV for the smoothed data in Fig. 3. 
The mean deviation for the AME2020 nuclides is less than 1 MeV and shows a clear dependence on $L$. Meanwhile, the rms deviation  for the AME2020 nuclides is about 3 MeV at $L \lesssim 90$ MeV but increases with $L$ clearly at $L > 90$ MeV.  These values are small as a semiclassical theory, which neglects shell energies of the order of MeV.

The mass deviations are significant for the MSIs, whose masses are lowered by the relatively large shell energies, with minor sensitivity to  $L$. Meanwhile, for unstable nuclei, the sensitivity to $L$ emerges with neutron-richness \cite{Oyamatsu:2010bf}, and the shell effects diminish. This is the origin of the $L$-sensitivity of the mean deviation. 
The AME1983 nuclides lay close to the MSIs.
The mean and rms deviations and their $L$ dependences for the AME1983 nuclides are between those for the MSIs and AME2020 nuclides. 
%
It is noteworthy that if we compare the rms deviations for the AME1983 and AME2020 nuclides, the progress of the mass evaluation in the last about 40 years reveals that the preferred value of $L$ would roughly lay between 20 and  90 MeV. However, the lower bound is not constrained well by the nuclear mass data of unstable nuclei. This range is consistent with the evaluations by Lattimer and Lim \cite{Lattimer:2012xj} and with recent experimental estimates of the $L$ value, which still have significant uncertainties. From the Sn+Sn reaction, Estee et al. estimate $42 < L < 117$ MeV \cite{SRIT:2021gcy}. From the parity violating asymmetry in $^{208}$Pb, Reed et al. estimate $L=106 \pm 37$ MeV \cite{Reed:2021nqk}, while Reinhard et al. analyze the same experimental data and evaluate $L=54 \pm 8$ MeV \cite{Reinhard:2021utv}. 

\clearpage

\section{Liquid-drop energies in the OI model}
\label{LiquidDrop_energies_in_OI_model}

\subsection{Incompressible liquid-drop mass formula}
\label{EOS_parameters_and_liquid_drop_mass_formula}

The proton, neutron, and mass numbers are continuous variables in this section.
The incompressible liquid drop (ILD) mass formula gives the mass excess as 
\begin{equation}
M_{ex\_ILD}= a_v A + a_I \frac{(N-Z)^2}{A}+ a_{surf} A^{2/3} + a_{C} \frac{Z^2}{A^{1/3}}   +\Delta m.
\label{MexLDref}
\end{equation}
The volume energy  $a_v$ is close to the saturation energy $w_0$ of symmetric nuclear matter. The surface energy coefficient $a_{surf}$ is  related to the gradient energy coefficient $F_0$ in Eq. (\ref{Wg_def}).   The symmetry energy coefficient $a_I$ is smaller than $S_0$ because it includes the energy of low-density matter at the surface. The Coulomb energy coefficient $a_{C}$ is related to nuclear size, hence to the saturation density $n_0$. 
The values of these four liquid drop coefficients are constrained well from nuclear masses. In this paper, we adopt the coefficient values of Yamada's reference ILD mass formula \cite{MYamada1964} in Table \ref{LDM_beta_stable_table}, which was determined from the overall fits of the $\beta$-stability line and the mass excesses of $\beta$-stable nuclei. Appendix \ref{comp_cal_exp} gives explicit formulae for neutron excess, mass, and radius of a nuclide on the smoothed $\beta$ stability line, and their calculated values with the coefficient values in Table \ref{LDM_beta_stable_table}.

\begin{table}[t]
\caption{The coefficient values in MeV of Yamada's reference ILD mass formula\cite{MYamada1964}.}
\begin{center}
\begin{tabular}{cccc}
\hline
$a_v$  & $a_C$ & $a_I$ & $a_s$\\
\hline
 -15.88485 & 0.71994 & 23.64332 & 18.32695 \\
\hline
\end{tabular}
\end{center}
\label{LDM_beta_stable_table}
\end{table}%

\begin{table}[t]
\caption{The energy parameters of the OI and ILD models, together with saturation parameters.}
\begin{center}
\begin{tabular}{ccccc}
\hline
  & symmetric matter & neutron matter & symmetry energy & finite range\\
\hline
interaction & isoscalar & isoscalar+isovector & isovector & isoscalar\\
OI & $a_1, \ a_2, \ a_3$ & $b_1, \ b_2(, \ b_3)$ & $a_1-a_3, \ b_1, \ b_2(, \ b_3)$ & $F_0$\\
ILD & $a_v, \ a_C$ & $a_v+a_I$ & $a_I$ & $a_s$\\
saturation & $n_0, \ w_0, \ K_0$ & $w_{n0}, \ L_{n0}$ & $S_0, \ L$ & \\
\hline
\end{tabular}
\end{center}
\label{parameter_correspondance_table}
\end{table}%

Table \ref{parameter_correspondance_table} summarizes the energy parameters in the OI and ILD models together with saturation parameters. 
For fixed $K_0$ and $L$, the OI model has the same degrees of freedom ($n_0, w_0, S_0$, and $F_0$) as the ILD model.
Consequently,
we can construct a family of the EOSs as a function of  $(K_0, L)$ by fitting nuclear masses. 
Furthermore, the five potential  parameters, $a_1-a_3, b_1$, and $b_2$, can be calculated analytically from the five saturation parameters,$n_0, w_0, K_0, S_0, L$, and $b_3$, as shown in Appendix \ref{interaction_parameters_from_saturation_parameters_b3}. Therefore, the interaction parameters are also functions of $(K_0, L)$.

\subsection{Mapping to liquid drop energies}
\label{mapping_to_LD_formula}

In this section, we show how the surface, symmetry, and volume energies of the ILD model are represented in the OI model.  If necessary, the subscripts "\_ILD" and "\_OI"  distinguish the models explicitly.

We use the size-equilibrium condition for the most stable nuclide to define the surface energy, $W_{surf\_OI}$ of the most stable nuclide. 
In the ILD model, from the condition that $M_{ex}/A$ is minimum with respect to $A$ keeping $(N-Z)/A$ constant, we obtain the relation,
\begin{equation}
W_{surf}=2 W_{C}
\label{WsLD2WCLD}
\end{equation}
between the surface energy $W_{surf}$  and the Coulomb energy $W_{C}$.
In the OI model,  from the size equilibrium condition for $M_{ex}/A$ (see Appendix \ref{size_equilibrium}), we obtain a similar relation between the gradient and Coulomb energies as
\begin{equation}
W_g=W_{C\_OI}.
\label{WgWC}
\end{equation}
Assuming the relation (\ref{WsLD2WCLD}) also in the OI model, we obtain the surface energy of the OI model using Eqs. (\ref{WgWC}) and (\ref{Wg_def}).
\begin{equation}
W_{surf\_OI} = 2W_g =  2  \int d^3 r F_0 |\nabla n(r)|^2.
\label{Wsurf}
\end{equation}
The liquid-drop symmetry energy of the OI model, $W_{i\_OI}$, is the isovector part of the uniform-matter energy $W_{EOS}$.
\begin{equation}
W_{i\_OI}=\int d^3 r \left [\epsilon_0 \left(n_n(r),n_p(r)\right) - \epsilon_0 \left(n(r)/2,n(r)/2\right) \right].
\label{Wi}
\end{equation}
Finally,  the volume energy of the OI model, $W_{v\_OI}$, is the remaining energy given by
\begin{equation}
W_{v\_OI}=W_{EOS} + W_g - W_{surf\_OI} - W_{i\_OI}  =\int d^3 r \left [\epsilon_0 \left(n(r)/2,n(r)/2\right) - F_0 |\nabla n(r)|^2  \right],
\label{Wv}
\end{equation}
which is desirable isoscalar energy. Equations (\ref{Wsurf}) and (\ref{Wv}) suggest that half of the surface energy comes from the EOS energy.

\begin{figure}[htbp]
\begin{center}
\includegraphics[]{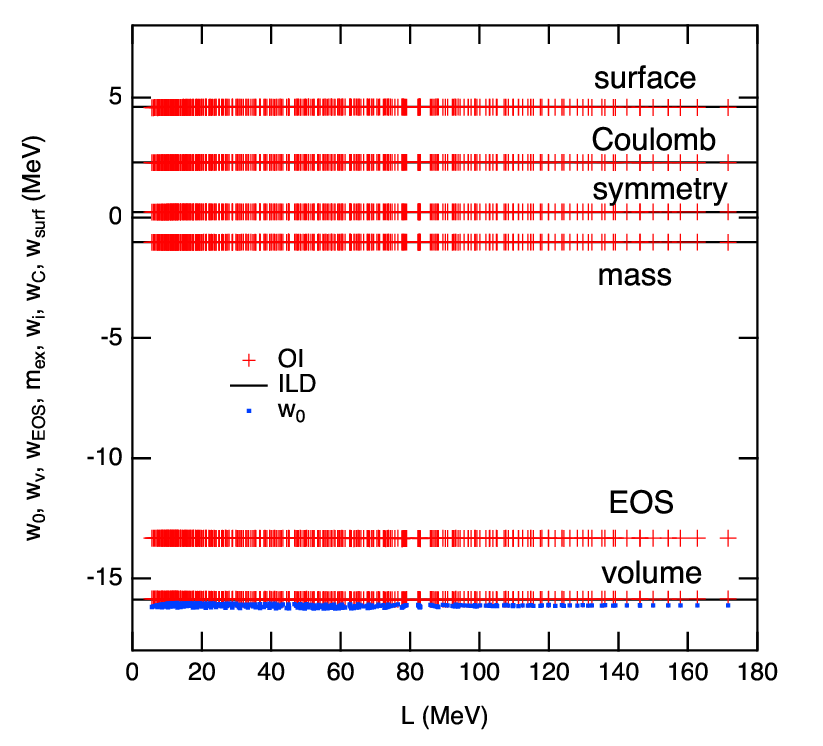}
\caption{The liquid-drop energies of the most stable nuclide as functions of $L$. The red crosses (OI) show OI model calculations, while the horizontal lines (ILD) are the results obtained with the coefficient values of the reference ILD model in Table \ref{LDM_beta_stable_table}. The blue dots ($w_0$) show the saturation energy $w_0$ of the OI model interactions.}
\label{fig_MSN_energy_contents}
\end{center}
\end{figure}

Figure \ref{fig_MSN_energy_contents} shows the OI and ILD energies per nucleon of the most stable nuclide; $w_{EOS}=W_{EOS}/A$, $w_{surf}=W_{surf}/A$, $w_C=W_C /A$, $w_i=W_i /A$, $w_v=W_v /A$,  and the mass excess per nucleon $m_{ex}=M_{ex}/A$.
These OI energies per nucleon are nearly constant of $L$ (and $K_0$) and almost equal to the ILD energies. We see that the OI model is a natural extension of the ILD model from the excellent agreement of the four liquid-drop energies,  the surface, Coulomb, symmetry, and volume energies.

As shown in Fig. \ref{fig_MSN_energy_contents},  the volume energy per nucleon $w_v$ is close to the saturation energy $w_0$ and surprisingly constant, presumably reflecting correlations among symmetric matter EOS (isoscalar) parameters, including $w_0 - F_0$. 
Hence,  the existence of the surface does not affect the liquid-drop core appreciably, thanks to the appropriate definitions of the surface and volume energies in Eqs. (\ref{Wsurf}) and (\ref{Wv}).



\begin{figure}[htbp]
\begin{center}
\includegraphics[]{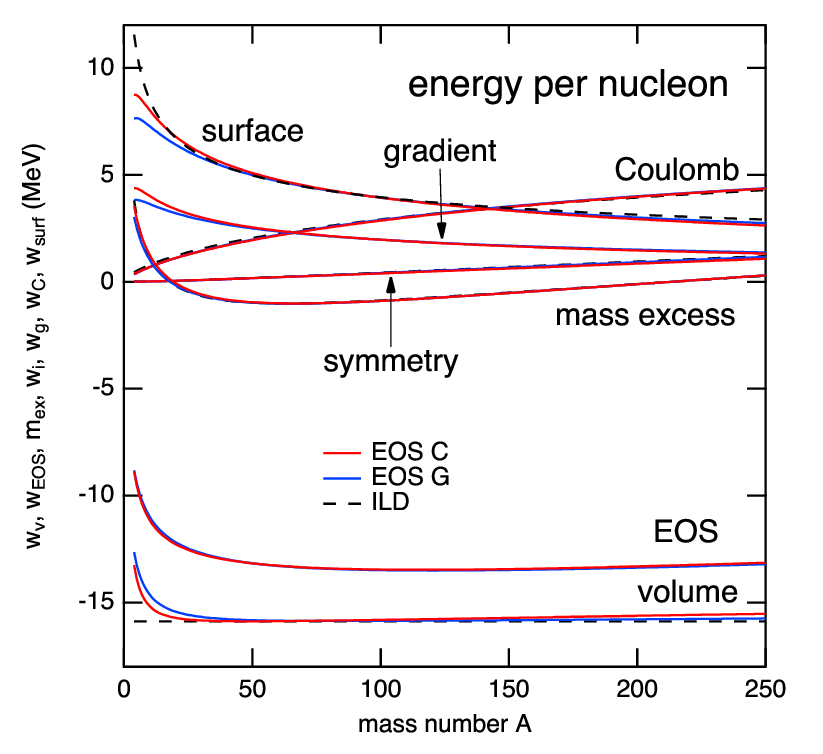}
\caption{The energies of the most stable isobars as functions of $A$. The red and blue lines show OI model calculations with two extreme EOSs C and G, while the black dashed lines are the results obtained with the reference ILD model. These three lines are almost indistinguishable except for the surface, gradient, and volume energies at small or large A values.}
\label{fig_w_energy_contents_A}
\end{center}
\end{figure}
\begin{figure}[htbp]
\begin{center}
\includegraphics[]{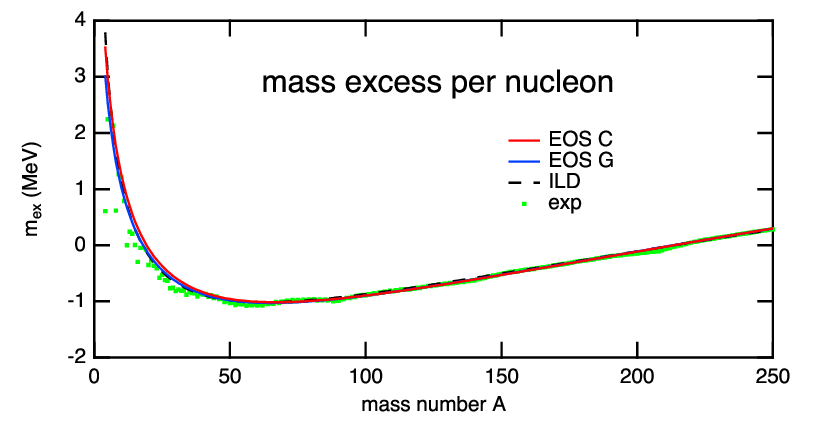}
\includegraphics[]{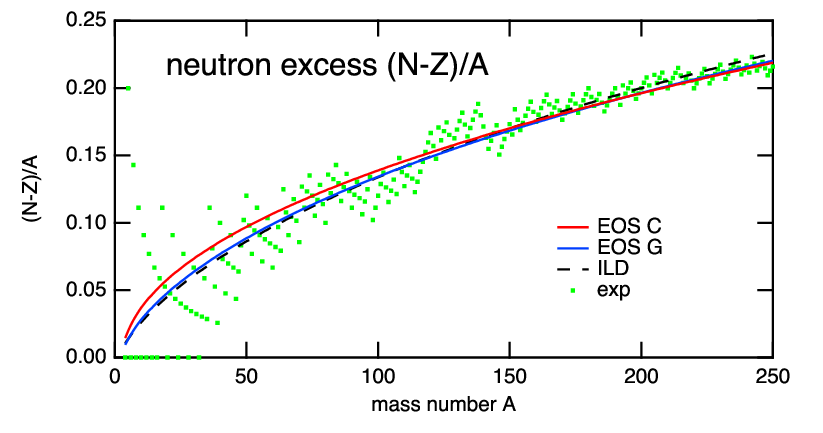}
\caption{The mass excess per nucleon  $m_{ex}$ and the neutron excess ratio $(N-Z)/A$ of the most stable isobars as functions of mass number $A$. The red and blue lines show the OI model calculations with two extreme EOSs C and G, while the black dashed lines are the reference ILD model calculations. The green dots show the experimental values.}
\label{fig_mex_alpha_MSI}
\end{center}
\end{figure}

Figure \ref{fig_w_energy_contents_A} shows that the OI model energies, even for two extreme EOSs C and G, also agree with the corresponding ILD energies as functions of mass number A. The agreement is excellent in the range of $A \gtrsim 40$.
Consequently, Fig. \ref{fig_mex_alpha_MSI} shows that the OI model values of the mass excess per nucleon $m_{ex}$ (upper) and the neutron excess ratio $(N-Z)/A$ (lower) also agree well with the experimental values.

Unfortunately, the OI and ILD models overestimate the mass excesses at $A \lesssim 40$, as shown in Fig. \ref{fig_mex_alpha_MSI} (upper). 
The increase of $w_{v\_OI}$ with the decrease of  $A$ implies that a better description of the surface energy is necessary.
The OI model's interaction and density distribution must be too crude to describe a light nucleus with $A \lesssim 25$, which has a small core compared to the surface and reduces its energy primarily by quantum mechanical effects.

Interestingly, the $(N-Z)/A$ values at $A \gtrsim 200$ of the two extreme OI models are almost equal but slightly different from that of the ILD. The neutron excess is mainly determined from the symmetry and Coulomb energies, and in the ILD model, it is given by
\begin{equation}
\frac{I}{A}=\frac{a_c A^{2/3}-(m_n-m_p-m_e)}{a_c  A^{2/3}+4a_{I}}.
\label{IbyA_MSI}
\end{equation}
In Fig. \ref{fig_mex_alpha_MSI}, the Coulomb and symmetry energies of the ILD and the OI models (EOSs C and G) agree well. In contrast, the surface and, consequently, volume energies (see Eq. (\ref{Wv})) of the EOSs C and G are slightly different from those of the ILD model. 
Hence, the slight neutron excess difference at $A \gtrsim 200$ between the ILD and OI models might be induced by the surface distribution difference (including the neutron skin) .
%

\clearpage

\subsection{Most stable nuclide in the OI and ILD models}
\label{MSI_OI_ILD}

\begin{figure}[htbp]
\begin{center}
\includegraphics[]{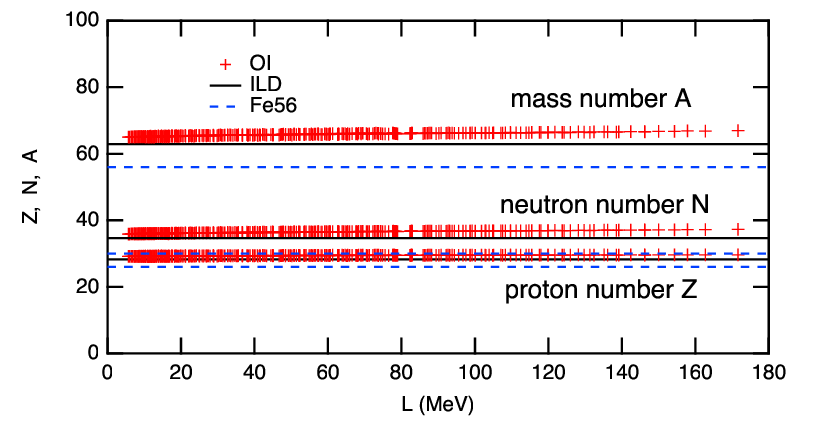}
\includegraphics[]{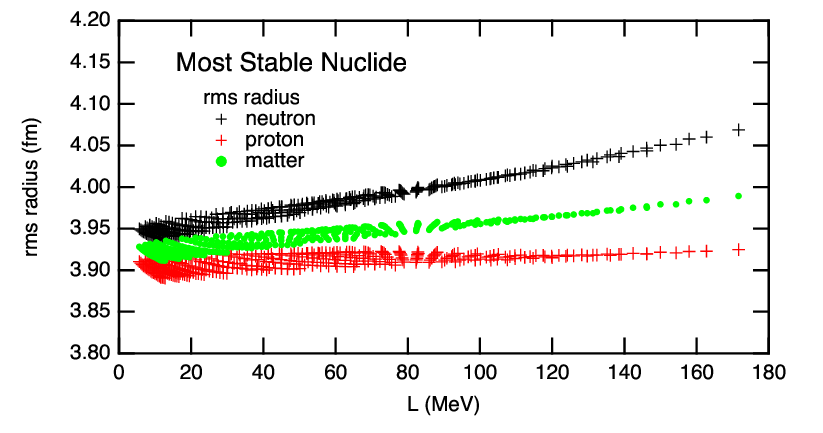}
\caption{The mass, neutron, and proton numbers (upper) and the rms radii (lower) of the most stable nuclide as functions of $L$. The crosses show the OI model calculations,  while the black horizontal lines are the results of the reference ILD model.}
\label{fig_MSN_location}
\end{center}
\end{figure}

Figure \ref{fig_MSN_location} (upper) shows that the most stable nuclides in the OI and ILD models are heavier than the empirical $^{56}$Fe ($Z=26$ and $A=56$). These deviations are not surprising because the shell energy significantly shifts the minimum point. In contrast, the smoothed energy per nucleon varies slowly around the minimum,  as shown in Fig. \ref{fig_mex_alpha_MSI}. Even for the KTUY mass formula \cite{Koura2005}, the gross part of  KTUY mass per nucleon is minimum at $Z=28$ and $A=62$.
Figure \ref{fig_MSN_location} (lower) shows that the proton, neutron, and matter radii are almost constant except for subtle $L$ dependence related to neutron skin formation. We note that the proton radius is almost constant of $K_0$ and $L$ because we fit the empirical charge radius data. 

\begin{figure}[htbp]
\begin{center}
\includegraphics[]{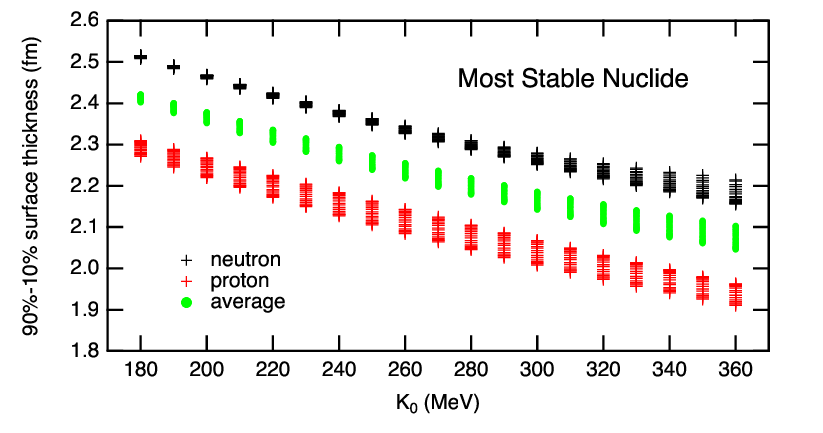}
\includegraphics[]{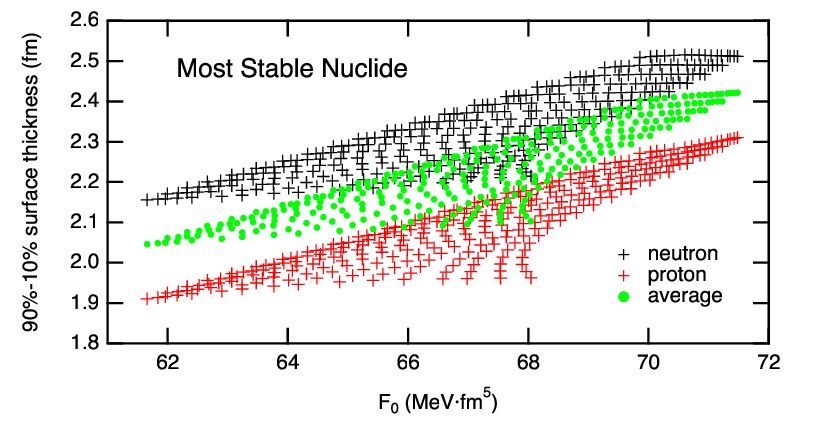}
\caption{The 90\%-10\% surface thicknesses of the point-neutron and point-proton distributions as functions of $K_0$ (upper) and $F_0$ (lower). Also shown is the average of these two thicknesses as a rough estimate of the matter distribution (Eq. (\ref{matter_rms_radius})).}
\label{fig_MSN_surface_thickness}
\end{center}
\end{figure}

Lastly, we mention that the point nucleon distribution (Eq. (\ref{density_parametrization})) of the most stable nucleus has a reasonable surface thickness, although we imposed no constraint on the thickness in optimizing the OI model parameters. 
Figure \ref{fig_MSN_surface_thickness} shows the 90\%-10\% surface thicknesses of the neutron and proton distributions denoted by thick(neutron) and thick(proton), together with the average of the two thicknesses as a rough estimate of the thickness of the matter distribution. 
\begin{equation}
\textrm{thick(average)} = \frac{N}{A} \textrm{thick(neutron)} + \frac{Z}{A} \textrm{thick(proton)}.
\label{matter_rms_radius}
\end{equation}
These three thicknesses correlate clearly with  $K_0$  (upper box) and appreciably with $F_0$  (lower box). 
These correlations reflect that $K_0$ and $F_0$ are the lowest-order parameters contributing to the surface energy.
While the surface energy agrees well between the OI and ILD models, the proton and average thickness values are slightly smaller than the empirical value of 2.4 fm in most cases \cite{OI2003}.

\clearpage

\section{Conclusions}
\label{conclusions}
This paper studies how nuclear masses are affected by the equation of state of nuclear matter. We adopt a macroscopic nuclear model, named the OI model, with reasonable many-body energy and isoscalar gradient energy. We use 304 update interactions, covering wide ranges of the incompressibility $K_0$ and the density-slope $L$.  For fixed $K_0$ and $L$, the OI model has the same number of independent interaction parameters as the ILD model. Moreover, all the OI interactions almost equally fit empirical mass, neutron excess, and radius data of stable nuclei,  nearly insensitively of $K_0$ and $L$. 

This insensitivity is consistent with the ILD picture and leads to the correlations among interaction and saturation parameters. 
We found that the interaction and saturation parameters of symmetric nuclear matter correlate mainly with  $K_0$, and those of neutron matter mainly with $L$. 

We assume that the surface energy of the OI model is twice as large as the gradient energy using the size equilibrium conditions of the ILD and OI models. Then, the two models'  volume, surface, symmetry, and Coulomb energies agree very well for the most stable isobars with $A \ge 40$. 

The correlation between the saturation energy $w_0$ and the gradient energy coefficient $F_0$ probably works to define the volume and surface energies properly.  Meanwhile, the well-known strong correlation between $S_0$ and $L$ helps explain the symmetry energy agreement between the ILD and OI models. Furthermore, the latter correlation 
causes the fixed points of the density-dependent symmetry energy and neutron matter EOS.

While calculated masses in the OI model are essentially insensitive to $K_0$ and $L$ for nuclei close to the $\beta$-stability line, they are relatively sensitive to $L$ for unstable nuclei. 
%
Interestingly, the OI model with $L \lesssim 100$ MeV predicts the latest mass data better than those of stable nuclei,  and we suggest $20 \lesssim  L \lesssim 90$ MeV, although the lower boundary is not constrained well.

Presumably, the conclusions of this paper are not affected significantly by our choice of the empirical data of stable nuclei, the many-body energy, or the inhomogeneity energy because the correlations among the saturation parameters of the OI model are consistent with relativistic and non-relativistic phenomenological interactions of contemporary use.

We have extensively performed the neutron star matter calculation with the OI model and have reported preliminary results in Refs. \cite{Oyamatsu:2017qzv} and \cite{ Oyamatsu:2020dlb}. We are preparing to publish the final results with discussions along the line of this paper.

\section*{Acknowledgment}

The author wishes to express sincere gratitude to the late Prof. M. Yamada for his thoughtful guidance on nuclear physics and the outstanding works on which the present paper relies. He also thanks Prof. K. Iida for his critical comments and discussions and Dr. H. Sotani for his basic but essential questions and comments.  
Finally, he also acknowledges this work is indebted to conversations with Profs. H. Toki, H. Shen, K. Sumiyoshi, H. Koura, K. Arita, K. Nakazato, Drs. H. Togashi, and A. Kohama.


%



\let\doi\relax


\begin{thebibliography}{99}



\bibitem{BohrMottelson1968} A. Bohr and B.R. Mottelson, Nuclear Structure I, (W. A. Benjamin,1968) p.143.

\bibitem{Lombard1973} R. J. Lombard, \ANN{77,380,1973}

\bibitem{OI2003} K. Oyamatsu and K. Iida, \PTP{109,631,2003}

\bibitem{OI2007} K. Oyamatsu and K. Iida, \PRC{75,015801,2007}
\bibitem{Oyamatsu:2010bf}
K.~Oyamatsu and K.~Iida,
\PRC{81,054302,2010}
\bibitem{Oyamatsu:2010sk}
K.~Oyamatsu, K.~Iida and H.~Koura,
\PRC{82,027301,2010}
\bibitem{Iida:2013fra}
K.~Iida and K.~Oyamatsu,
Eur. Phys. J. A, \textbf{50}, 42 (2014).
\bibitem{Sotani:2013dga}
H.~Sotani, K.~Iida, K.~Oyamatsu and A.~Ohnishi,
PTEP, \textbf{2014}, 051E01 (2014).
\bibitem{Sotani:2012qc}
H.~Sotani, K.~Nakazato, K.~Iida and K.~Oyamatsu,
\PRL{108,201101,2012}
\bibitem{Sotani:2012xd}
H.~Sotani, K.~Nakazato, K.~Iida and K.~Oyamatsu,
Mon. Not. Roy. Astron. Soc., \textbf{428}, L21 (2013).
\bibitem{Sotani:2013jya}
H.~Sotani, K.~Nakazato, K.~Iida and K.~Oyamatsu,
Mon. Not. Roy. Astron. Soc., \textbf{434}, 2060 (2013).
\bibitem{Sotani:2015lya}
H.~Sotani, K.~Iida and K.~Oyamatsu,
\PRC{91,015805,2015}
\bibitem{Sotani:2015laa}
H.~Sotani, K.~Iida and K.~Oyamatsu,
New Astron., \textbf{43}, 80 (2016).
\bibitem{Sotani:2016pmb}
H.~Sotani, K.~Iida and K.~Oyamatsu,
Mon. Not. Roy. Astron. Soc., \textbf{464}, 3101 (2017).
\bibitem{Sotani:2017hpq}
H.~Sotani, K.~Iida and K.~Oyamatsu,
Mon. Not. Roy. Astron. Soc., \textbf{470}, 4397 (2017).
\bibitem{Sotani:2018tdr}
H.~Sotani, K.~Iida and K.~Oyamatsu,
Mon. Not. Roy. Astron. Soc., \textbf{479}, 4735 (2018).
\bibitem{Sotani:2019pja}
H.~Sotani, K.~Iida and K.~Oyamatsu,
Mon. Not. Roy. Astron. Soc., \textbf{489}, 3022 (2019).


\bibitem{Hohenberg:1964zz}
P.~Hohenberg and W.~Kohn,
Phys. Rev. \textbf{136}, B864-B871 (1964).

\bibitem{Kohn:1965zza}
W.~Kohn and L.~J.~Sham,
Phys. Rev. \textbf{137}, A1697-A1705 (1965).


\bibitem{BrackGuetHakansson1985} M. Brack, C. Guet and H.-B. Hakansson, \PRP{123,273,1985}


\bibitem{MYamada1964} M. Yamada, 
\PTP{32,512,1964}

\bibitem{YamadaMatumoto1961JPSJ} M. Yamada and Z. Matumoto, \JPSJ{16, 1497,1961}



\bibitem{1961NuclidicMassTable} L.~A. ~K\"onig, J.~H.~E. ~Mattauch, A.~H. ~Wapstra,
Nucl. Phys. \textbf{31},18-42 (1962).



\bibitem{Reed:2021nqk}
B.~T.~Reed, F.~J.~Fattoyev, C.~J.~Horowitz and J.~Piekarewicz,
Phys. Rev. Lett. \textbf{126}, no.17, 172503 (2021).

\bibitem{Reinhard:2021utv}
P.~G.~Reinhard, X.~Roca-Maza and W.~Nazarewicz,
Phys. Rev. Lett. \textbf{127}, no.23, 232501 (2021).



\bibitem{SRIT:2021gcy}
J.~Estee \textit{et al.} [S$\pi$RIT],
Phys. Rev. Lett. \textbf{126}, no.16, 162701 (2021).


\bibitem{oya1993} K. Oyamatsu, \NPA{561,431,1993}

\bibitem{BludmanDover1980} S. A. Bludman and C. B. Dover, 
\PRD{22,1333,1980}
\bibitem{FriedmanPandharipande1980} B. Friedman and V.R. Pandharipande, 
\NPA{361,502,1981}

\bibitem{EltonSwift67} L. R. B. Elton and A. Swift, 
\NPA{94,52,1967}

\bibitem{1987ChargeRadii} H. de Vries, C. W. de Jager and C. de Vries, Atomic Data and Nuclear Data Tables, \textbf{36}, 251(1987).


\bibitem{KodamaYamada1971} T. Kodama and M. Yamada, 
\PTP{45,1763,1971}


\bibitem{TewsLattimerOhnishiKolomeitsev2018} I. Tews, J. M. Lattimer, A. Ohnishi and E. E Kolomeitsev, 
\AJ{848,105,2018}


\bibitem{ABrown2000} B. A. Brown, 
\PRL{85,5296,2000}

\bibitem{Brueckner:1968zzb}
K.~A.~Brueckner, J.~R.~Buchler, S.~Jorna and R.~J.~Lombard,
\PR{171,1188,1968}




\bibitem{Wang:2021xhn}
M.~Wang, W.~J.~Huang, F.~G.~Kondev, G.~Audi and S.~Naimi,
Chin. Phys. C, \textbf{45}, 030003 (2021).

\bibitem{Wapstra:1985hmr}
A.~H.~Wapstra and G.~Audi,
\NPA{432,55,1985}


\bibitem{Lattimer:2012xj}
J.~M.~Lattimer and Y.~Lim,
Astrophys. J. \textbf{771}, 51 (2013).


\bibitem{Koura2005}
H.~Koura, T.~Tachibana, M.~Uno and M.~Yamada, 
\PTP{113,305,2005}

\bibitem{Oyamatsu:2017qzv}
K.~Oyamatsu, H.~Sotani and K.~Iida,
PoS \textbf{INPC2016}, 136 (2017).

\bibitem{Oyamatsu:2020dlb}
K.~Oyamatsu, K.~Iida and H.~Sotani,
J. Phys. Conf. Ser. \textbf{1643}, no.1, 012059 (2020).


\bibitem{ARPONEN1972257} J. Arponen, 
\NPA{191,257,1972}




\end{thebibliography}

\appendix

\section{Explicit formula of saturation parameters and potential parameters}
\label{explicit_formula_of_saturation_parameters_w_potential_parameters}
Our previous paper \cite{OI2003} simplified the kinetic energy expression by using the neutron mass as the proton mass because this replacement gives a relatively small effect. Meanwhile, the numerical calculations in the paper were performed using the real proton mass. In this paper, we write the neutron and proton rest masses explicitly.

The  neutron-proton mass difference forces us to use some prescription 
for the definition of the density-dependent symmetry energy $S(n)$
because the kinetic energy  density in $\epsilon_0(n_n, n_p)$ (Eq. (\ref{homogineoues_energy_density})) is not  charge-symmetric, and
\begin{equation}
\frac{\partial w(n,\alpha)}{\partial \alpha}\bigg|_{\alpha=0} \ne 0.
\end{equation}
To keep the charge symmetry,  we use the value of the nucleon rest mass of 
\begin{equation}
m=\frac{m_n+m_p}{2},
\end{equation}
when and only when we calculate $S(n)$ and its saturation parameters. This prescription increases the kinetic energy by $5\times10^{-5}$\%. 

\subsection{Symmetric matter EOS $w_s(n)$}
\label{symmetric_matter_eos}
The proton rest mass $m_p$ and the neutron rest mass $m_n$  are used.
\begin{equation}
w_s(n) = c_s n^{2/3} + a_1 n + \frac{a_2 n^2}{1+a_3 n}, 
\label{ws_n_def}
\end{equation}
with
\begin{equation}
c_s= \frac{3}{10}\left(\frac{3\pi^2}{2}\right)^{2/3}\left(\frac{\hbar^2}{2m_n}+\frac{\hbar^2}{2m_p}\right).
\end{equation}
\begin{equation}
w_0 = c_s n_0^{2/3} + a_1 n_0 + \frac{a_2 n_0^2}{1+a_3 n_0}.
\label{w0_def}
\end{equation}
\begin{equation}
L_{0}=3 n_0 \frac{dw_s}{dn}\Big|_{n=n_0} = 2 c_s n_0^{2/3} + 3 a_1 n_0 + \frac{3 a_2 n_0^2 (2+a_3 n_0)}{(1+a_3 n_0)^2}=0.
\label{L0_def}
\end{equation}
\begin{equation}
K_{0}=9 n_0^2 \frac{d^2w_s}{d n^2}\Big|_{n=n_0} = -2  c_s n_0^{2/3} + \frac{18 a_2 n_0^2 }{(1+a_3 n_0)^3}.
\label{K0_def}
\end{equation}
\begin{equation}
Q_{0}=27 n_0^3 \frac{d^3w_s}{d n^3}\Big|_{n=n_0} = 8  c_s n_0^{2/3} - \frac{162 a_2 a_3 n_0^3}{(1+a_3 n_0)^4}.
\label{Q0_def}
\end{equation}

\subsection{Neutron matter EOS $w_n(n)$}
\label{neutron_matter_eos}
The neutron rest mass $m_n$ is used.
\begin{equation}
w_n(n)=c_n n^{2/3} + b_1 n + \frac{b_2 n^2}{1+b_3 n}, \quad c_n = \frac{3}{5}(3\pi^2)^{2/3}\frac{\hbar}{2m_n}, 
\label{wn_n_def}
\end{equation}
with
\begin{equation}
c_n = \frac{3}{5}(3\pi^2)^{2/3}\frac{\hbar^2}{2m_n}.
\end{equation}
\begin{equation}
w_{n0} = c_n n_0^{2/3} + b_1 n_0 + \frac{b_2 n_0^2}{1+b_3 n_0}.
\label{wn0_def}
\end{equation}
\begin{equation}
L_{n0}=3 n_0 \frac{dw_n}{dn}\Big|_{n=n_0} = 2 c_n n_0^{2/3} + 3 b_1 n_0 + \frac{3 b_2 n_0^2 (2+b_3 n_0)}{(1+b_3 n_0)^2}.
\label{Ln0_def}
\end{equation}
\begin{equation}
K_{n0}=9 n_0^2 \frac{d^2w_n}{d n^2}\Big|_{n=n_0} = -2 c_n n_0^{2/3} + \frac{18 b_2 n_0^2}{(1+b_3 n_0)^3}.
\label{Kn0_def}
\end{equation}
\begin{equation}
Q_{n0}=27 n_0^3 \frac{d^3w_n}{d n^3}\Big|_{n=n_0} = 8 c_n n_0^{2/3} - \frac{162 b_2 b_3 n_0^3}{(1+b_3 n_0)^4}.
\label{Qn0_def}
\end{equation}

\subsection{Density-dependent symmetry energy $S(n)$}
\label{density_dependent_symmetry_energy}
The nucleon mass $m=(m_n+m_p)/2$ is used as described at the beginning of this Appendix.
\begin{equation}
S(n)=\frac{\partial w}{\partial \alpha^2}\Big|_{\alpha=0}=c_m n^{2/3} +  (b_1 - a_1) n + \frac{b_2 n^2}{1+b_3 n} - \frac{a_2 n^2}{1+a_3 n},
\label{S_n_def}
\end{equation}
with
\begin{equation}
c_m= \frac{1}{3}\left(\frac{3\pi^2}{2}\right)^{2/3}\frac{\hbar^2}{2m}.
\end{equation}
\begin{equation}
S_0= c_m n_0^{2/3} + (b_1 - a_1) n_0 + \frac{b_2 n_0^2}{1+b_3 n_0} - \frac{a_2 n_0^2}{1+a_3 n_0}.
\label{S0_def}
\end{equation}
\begin{equation}
L=3 n_0 \frac{dS}{dn}\Big|_{n=n_0} = 2 c_m n_0^{2/3} + 3(b_1 - a_1) n_0 + \frac{3 b_2 n_0^2 (2+b_3 n_0)}{(1+b_3 n_0)^2} - \frac{3 a_2 n_0^2 (2+a_3 n_0)}{(1+a_3 n_0)^2}.
\label{L_def}
\end{equation}
\begin{equation}
K_{sym}=9 n_0^2 \frac{d^2 S}{d n^2}\Big|_{n=n_0} = -2 c_m n_0^{2/3}  + \frac{18 b_2 n_0^2}{(1+b_3 n_0)^3} - \frac{18a_2 n_0^2}{(1+a_3 n_0)^3}.
\label{Ksym_def}
\end{equation}
\begin{equation}
Q_{sym}=27 n_0^3 \frac{d^3 S}{d n^3}\Big|_{n=n_0} = 4 c_m n_0^{2/3}  - \frac{162 b_2 b_3 n_0^3}{(1+b_3 n_0)^4} + \frac{162 a_2 a_3 n_0^3}{(1+a_3 n_0)^4}.
\label{Qsym_def}
\end{equation}

\subsection{Potential parameters $a_1, a_2, a_3, b_1$ and $b_2$}
\label{interaction_parameters_from_saturation_parameters_b3}
We calculate the potential parameters $a_1, a_2, a_3, b_1$, and $b_2$ from the saturation parameters $n_0, w_0, K_0, S_0$, and $L$,  and the potential parameter $b_3$ using the following formula.
\begin{equation}
a_3=\frac{1}{n_0} \frac{4 c_s n_0^{2/3} -18 w_0 - K_0}{2 c_s n_0^{2/3} + K_0} .
\label{a3_value}
\end{equation}
\begin{equation}
a_2=\frac{(1+a_3 n_0)^3}{18 n_0^2} \left(2 c_s n_0^{2/3} + K_0 \right) .
\label{a2_value}
\end{equation}
\begin{equation}
a_1=\frac{1}{n_0} \left(w_0 - c_s n_0^{2/3} - \frac{a_2 n_0^2}{1+a_3 n_0} \right) .
\label{a1_value}
\end{equation}
\begin{equation}
b_2=\frac{(1+b_3 n_0)^2}{3 n_0^2} \left[\frac{5}{9} c_m n_0^{2/3} + \frac{3 a_2 n_0^2}{(1+a_3 n_0)^2} - 3 S_0 + L \right].
\label{b2_value}
\end{equation}
\begin{equation}
b_1=\frac{1}{n_0} \left( - \frac{5}{9} c_m n_0^{2/3} + a_1 n_0  + \frac{a_2 n_0^2}{1+a_3 n_0} 
 -  \frac{b_2 n_0^2}{1+b_3 n_0} + S_0 \right) .
\label{b1_value}
\end{equation}

\section{Comparison between the calculated and empirical values of $I$, $M_{ex}$, and $R_{ch}$}
\label{comp_cal_exp}

\begin{figure}[htbp]
\begin{center}
\includegraphics[width=8cm]{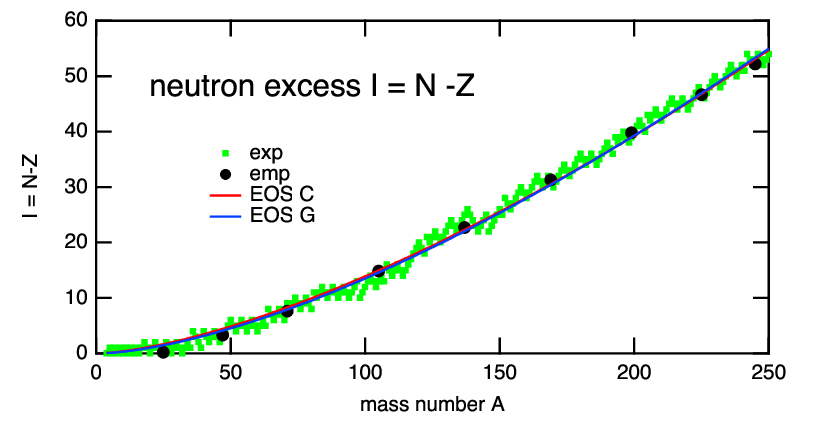}
\includegraphics[width=8cm]{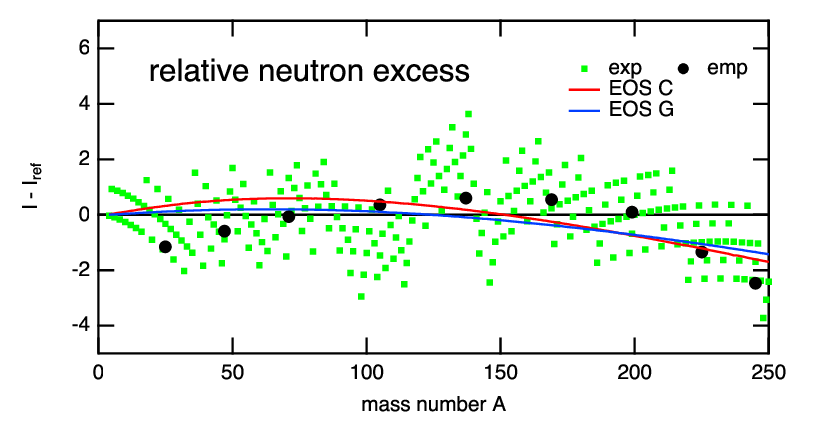}
\includegraphics[width=8cm]{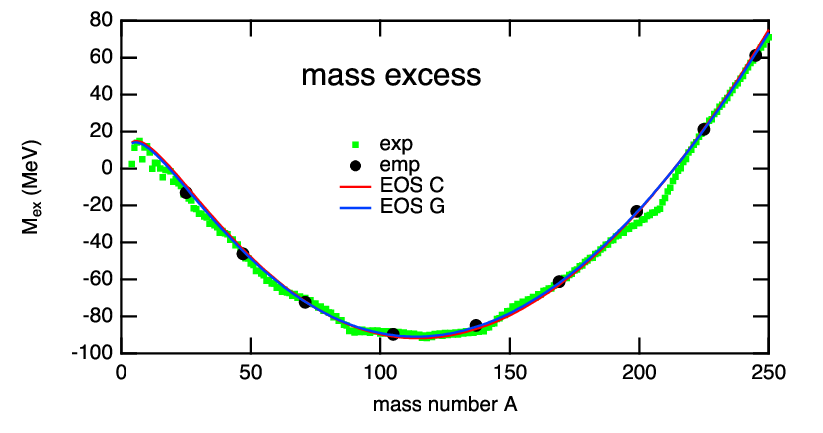}
\includegraphics[width=8cm]{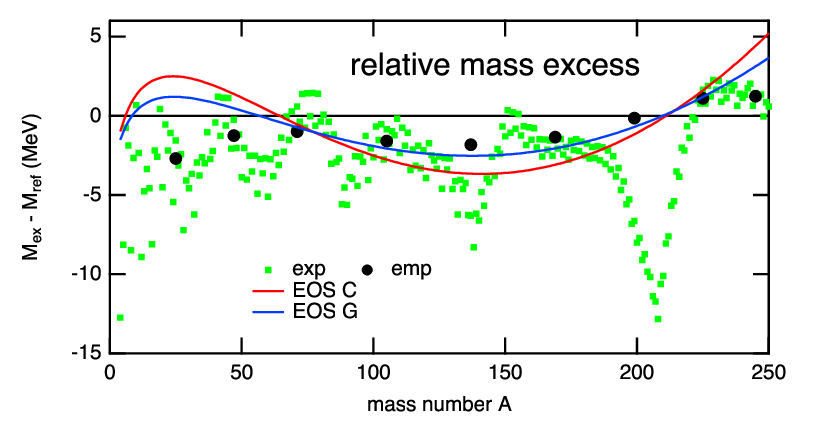}
\includegraphics[width=8cm]{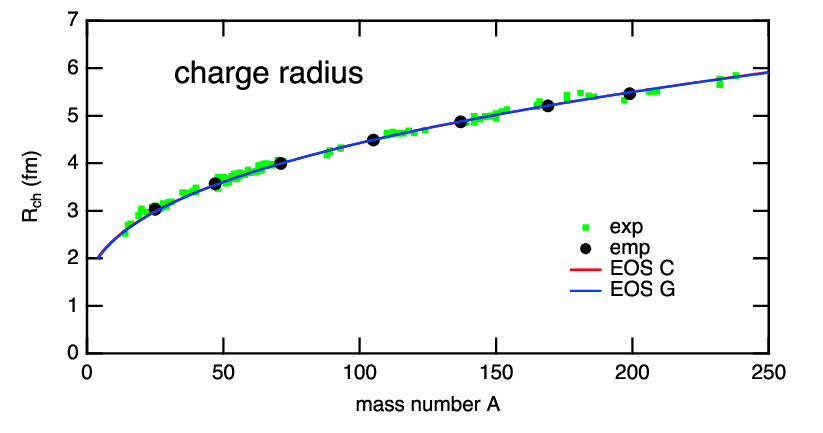}
\includegraphics[width=8cm]{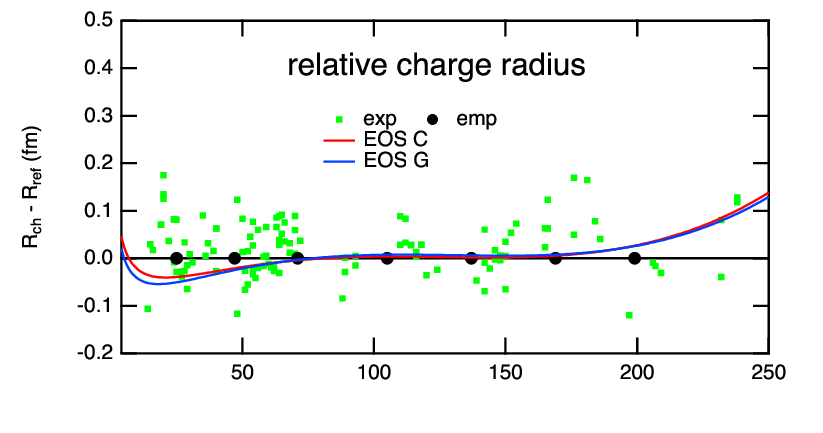}
\caption{Comparison of the empirical (filled black circles) and calculated (lines) values of  $I$, $M_{ex}$, and $R_{ch}$.  The calculations are performed with two extreme EOSs C (red lines) and G (blue lines) in Table \ref{EOS_ACGI_oya93}. For $I$ and  $M_{ex}$, the experimental values of the most stable isobars \cite{Wang:2021xhn} are also shown by green dots. The experimental $R_{ch}$ values (green dots) are taken from the 1987 compilation of charge radii. \cite{1987ChargeRadii}. }
\label{fit_I_Mex_Rch}
\end{center}
\end{figure}

Figure \ref{fit_I_Mex_Rch} compares the calculated and empirical values of  $I$, $M_{ex}$, and $R_{ch}$. The deviations from the empirical values are small compared with the scatterings due to shell effects even for two extreme EOSs C and G.

To compare the empirical and calculated values in detail in Fig. \ref{fit_I_Mex_Rch}, we use the following reference formulae: approximate smoothed values of the neutron excess $I_{ref}$
\cite{KodamaYamada1971}
, the mass excess $M_{ref}$
\cite{MYamada1964, KodamaYamada1971}
, and the charge radius $R_{ref}$ as functions of mass number $A$. 
\begin{equation}
I_{ref}=\frac{0.35997 A^{2/3}-0.39131 }{0.35997 A^{2/3}+47.28664} \ A
\label{I_ref}
\end{equation}
\begin{equation}
\begin{split}
M_{ref} =(7.68004-15.88485)A+0.39131 I_{ref}+18.32695 A^{2/3}\\
+23.64332 \frac{I_{ref}^2}{A}+0.71994 \frac{(A-I_{ref})^2}{4 A^{1/3}}
\end{split}
\label{Mex_ref}
\end{equation}
\begin{equation}
\begin{split}
R_{ref}=0.665608+0.830616 A^{1/3}-3.3634\times 10^{-3} A\\
+3.1635\times 10^{-5} A^2-8.0277\times 10^{-8} A^3
\end{split}
\label{Rch_ref}
\end{equation}
Equation (\ref{Rch_ref}) was also used to calculate the empirical $R_{ch}^{emp}$ values in Table \ref{empirical_data_tbl}.

\clearpage

\section{Inhomogeneity energy and symmetry energy $S_0$ in our early  study}
\label{notes_on_oya1993}
The inhomogeneity energy is the sum of the kinetic and potential contributions\cite{Brueckner:1968zzb}. The latter comes from the long-range part of the internucleon interactions\cite{Brueckner:1968zzb}. The OI model approximates the inhomogeneity energy density by the isoscalar gradient energy density;
\begin{equation}
\epsilon_g  \left(n_n,n_p\right) = F_0|\nabla n(r)|^2.
\label{epsilon_g_def}
\end{equation}
The value of the empirical parameter $F_0$ effectively includes both kinetic and potential contributions.

In our early study of neutron star matter \cite{oya1993}, the present author used the inhomogeneity energy density of the lowest order
\begin{equation}
\begin{split}
 \epsilon_g \left(n_n, n_p; \nabla n_n, \nabla n_p \right)=
 \frac{\alpha}{36} \left[\frac{\hbar^2}{2m_n}\frac{\nabla^2 n_n(r)}{n_n(r)}+\frac{\hbar^2}{2m_p}\frac{\nabla^2 n_p(r)}{n_p(r)}\right]  \\
 +  F_0 \left[ |\nabla n(r)|^2 - \beta |\nabla n_n(r) - \nabla n_p(r)|^2\right]
\end{split}
\label{oya1993_grad_energy_density}
\end{equation}
with $\alpha, \beta =0,1$. The values of $\alpha$ and $\beta$ of the current OI and early oya1-4 \cite{oya1993} models are listed in Table \ref{parameter_alpha_beta_values}.
The first term of the right-hand side of Eq. (\ref{oya1993_grad_energy_density}) is the kinetic energy density, while the term with coefficient $\beta$ is the isovector potential energy density.
However, $\alpha=1$ or $\beta=1$ makes little difference either in stable laboratory nuclei or inner-crust nuclei \cite{oya1993}.
It is also noted that the functional form of the density distribution (Eq. (\ref{density_parametrization})) was so modified, from Arponen's function \cite{ARPONEN1972257}, that the gradient energy density (Eq. (\ref{oya1993_grad_energy_density})) is continuous at $r=R_i$.

\begin{table}[h]
\caption{The $\alpha$ and $\beta$ values in Eq. (\ref{oya1993_grad_energy_density}) of the OI and oya1-4 \cite{oya1993} models.}
\begin{center}
\begin{tabular}{ccccc}
\hline
parameter  & oya1 & oya2 & oya3 & oya4 and OI\\
\hline
$\alpha$ & 1 & 1 & 1 & 0\\
$\beta$ & 0 & 1 & 0 & 0\\
\hline
\end{tabular}
\end{center}
\label{parameter_alpha_beta_values}
\end{table}%

The definition of the symmetry energy $S_0$ in Ref. \cite{oya1993} was
\begin{equation}
S_0=w_n(n_0)-w_s(n_0) =w_{n0} - w_0.
\label{S0_oya1993}
\end{equation}
In this paper, $S_0$ is defined from the density-dependent symmetry energy $S(n)=S^{(2)}(n)$;
\begin{equation}
S_0=S^{(2)}(n_0)= \frac{\partial w}{\partial \alpha^{2}}\Big|_{\alpha=0, n=n_0}.
\label{S0_present}
\end{equation}
The explicit formula of $S_0$ is given as a function of the potential parameters $a_1 -a_3, b_1 -b_3$ in Eq. (\ref{S0_def}).

\section{Size equilibrium conditions}
\label{size_equilibrium}

In the ILD mass formula, we take A and $\alpha=(N-Z)/A$ as independent variables so that 
\begin{equation}
W{_{C\_ILD}}=a_c \left(\frac{1-\alpha}{2}\right)^2 A^{5/3}.
\label{WC_A_alpha}
\end{equation}
Then, the size equilibrium condition is
\begin{equation}
\begin{split}
\frac{\partial}{\partial A} \frac{M_{ex}}{A}\Big|_\alpha=-\frac{1}{3} a_{surf} A^{-4/3} + \frac{2}{3} a_c \left(\frac{1-\alpha}{2}\right)^2 A^{-1/3}\\
=\frac{1}{3A^2}\left(-W_{surf\_ILD}+2W{_{C\_ILD}}\right)=0. 
\end{split}
\label{ILD_size_equi}
\end{equation}
We obtain the size equilibrium condition (Eq. (\ref{WsLD2WCLD})) from Eq. (\ref{ILD_size_equi}) for the ILD model.

In the OI model, we introduce a scale parameter $R$
in Eqs. (\ref{N_def})-(\ref{WC_def})  such that $\mathbf{r}=R \mathbf{u}$, $\tilde n(u) = n(r)$ and  $\tilde n_p(u) = n_p(r)$.
Then, the size equilibrium condition is 
\begin{equation}
\begin{split}
\frac{\partial}{\partial R} \frac{M_{ex}}{A}\Big|_R=\left(-\frac{2}{R^3} \int d^3 u F_0 |\nabla_u \tilde n|^2 +2 R \ \frac{e^2}{2} \int d^3 u d^3 u' \frac{\tilde n_p(u) \tilde n_p(u')}{|\mathbf{u} - \mathbf{u'}|} \right)/  \int d\mathbf{u} \tilde n(u)\\
=2 (RA)^{-1} (-W_g+W_{C\_OI})=0.
\end{split}
\label{OI_size_equi}
\end{equation}
We obtain the OI model condition (Eq. (\ref{WgWC})) from Eq. (\ref{OI_size_equi}). 


\section{Optimum values of saturation parameters and $F_0$}
Table \ref{OI_saturation_parameter_table} gives the updated optimal values of $n_0$, $w_0$,  $S_0$, and $F_0$ for the 304 sets of the $-y$ and $K_0$ values and corresponds to Table II in Appendix A of the previous study \cite{OI2003}. The $L$ values were calculated from $-y$, $K_0$, $n_0$, $w_0$, and $S_0$ using Eq. (\ref{slope_y}).

\begin{longtable}{ccccccc}
\caption{The updated optimal parameter values for the 304 sets of  the $-y$ and $K_0$ values.} 
\label{OI_saturation_parameter_table}\\
\hline
$-y$	& $K_0$	& $n_0$ &	 $w_0$ & $S_0$ & $L$ & $F_0$ \\
(MeV $\cdot$ fm$^3$)   &  (MeV)    & (MeV)  & (fm$^{-3}$)  &  (MeV)   & (MeV)  & (MeV $\cdot$  fm$^5$) \\
\hline
\endfirsthead

 & & & Table \ref{OI_saturation_parameter_table}\\
\hline
$-y$	& $K_0$	& $n_0$ &	 $w_0$ & $S_0$ & $L$ & $F_0$ \\
\hline
\endhead

\hline
\multicolumn{7}{r}{(continued)} \\ 
\endfoot


200	&	180	&	0.16928	&	-16.259	&	32.974	&	58.437	&	71.481	\\
200	&	190	&	0.16695	&	-16.245	&	33.236	&	63.042	&	71.228	\\
200	&	200	&	0.16485	&	-16.233	&	33.521	&	67.780	&	70.940	\\
200	&	210	&	0.16286	&	-16.220	&	33.799	&	72.636	&	70.646	\\
200	&	220	&	0.16117	&	-16.211	&	34.129	&	77.646	&	70.340	\\
200	&	230	&	0.15954	&	-16.200	&	34.458	&	82.792	&	70.025	\\
200	&	240	&	0.15787	&	-16.186	&	34.794	&	88.159	&	69.709	\\
200	&	250	&	0.15645	&	-16.178	&	35.230	&	93.828	&	69.387	\\
200	&	260	&	0.15526	&	-16.169	&	35.556	&	99.237	&	69.085	\\
200	&	270	&	0.15407	&	-16.163	&	35.985	&	105.11	&	68.798	\\
200	&	280	&	0.15293	&	-16.157	&	36.447	&	111.22	&	68.532	\\
200	&	290	&	0.15184	&	-16.151	&	36.920	&	117.52	&	68.318	\\
200	&	300	&	0.15071	&	-16.145	&	37.451	&	124.25	&	68.169	\\
200	&	310	&	0.14963	&	-16.139	&	38.022	&	131.28	&	68.023	\\
200	&	320	&	0.14868	&	-16.136	&	38.610	&	138.50	&	67.945	\\
200	&	330	&	0.14770	&	-16.132	&	39.273	&	146.24	&	67.903	\\
200	&	340	&	0.14677	&	-16.130	&	39.977	&	154.35	&	67.899	\\
200	&	350	&	0.14588	&	-16.129	&	40.705	&	162.77	&	67.963	\\
200	&	360	&	0.14495	&	-16.130	&	41.462	&	171.63	&	68.044	\\
210	&	180	&	0.16931	&	-16.258	&	32.687	&	55.160	&	71.422	\\
210	&	190	&	0.16715	&	-16.249	&	32.971	&	59.490	&	71.147	\\
210	&	200	&	0.16507	&	-16.236	&	33.197	&	63.842	&	70.855	\\
210	&	210	&	0.16316	&	-16.224	&	33.457	&	68.354	&	70.523	\\
210	&	220	&	0.16141	&	-16.212	&	33.737	&	72.990	&	70.195	\\
210	&	230	&	0.15978	&	-16.201	&	34.035	&	77.765	&	69.866	\\
210	&	240	&	0.15823	&	-16.192	&	34.466	&	82.978	&	69.542	\\
210	&	250	&	0.15679	&	-16.178	&	34.700	&	87.827	&	69.207	\\
210	&	260	&	0.15555	&	-16.173	&	35.035	&	92.953	&	68.919	\\
210	&	270	&	0.15420	&	-16.164	&	35.517	&	98.712	&	68.611	\\
210	&	280	&	0.15325	&	-16.159	&	35.804	&	103.84	&	68.342	\\
210	&	290	&	0.15213	&	-16.153	&	36.237	&	109.65	&	68.111	\\
210	&	300	&	0.15112	&	-16.148	&	36.659	&	115.52	&	67.925	\\
210	&	310	&	0.15011	&	-16.144	&	37.162	&	121.82	&	67.775	\\
210	&	320	&	0.14912	&	-16.136	&	37.640	&	128.21	&	67.654	\\
210	&	330	&	0.14807	&	-16.132	&	38.227	&	135.23	&	67.575	\\
210	&	340	&	0.14721	&	-16.131	&	38.843	&	142.40	&	67.538	\\
210	&	350	&	0.14630	&	-16.128	&	39.500	&	150.00	&	67.534	\\
210	&	360	&	0.14550	&	-16.129	&	40.205	&	157.90	&	67.538	\\
220	&	180	&	0.16921	&	-16.252	&	32.427	&	52.266	&	71.360	\\
220	&	190	&	0.16699	&	-16.240	&	32.648	&	56.284	&	71.063	\\
220	&	200	&	0.16490	&	-16.227	&	32.879	&	60.419	&	70.739	\\
220	&	210	&	0.16301	&	-16.215	&	33.123	&	64.652	&	70.410	\\
220	&	220	&	0.16130	&	-16.205	&	33.395	&	69.013	&	70.063	\\
220	&	230	&	0.15967	&	-16.193	&	33.643	&	73.425	&	69.711	\\
220	&	240	&	0.15819	&	-16.182	&	33.926	&	77.989	&	69.359	\\
220	&	250	&	0.15679	&	-16.172	&	34.222	&	82.677	&	69.014	\\
220	&	260	&	0.15549	&	-16.163	&	34.525	&	87.471	&	68.699	\\
220	&	270	&	0.15427	&	-16.155	&	34.864	&	92.453	&	68.396	\\
220	&	280	&	0.15319	&	-16.149	&	35.225	&	97.552	&	68.110	\\
220	&	290	&	0.15217	&	-16.144	&	35.597	&	102.79	&	67.841	\\
220	&	300	&	0.15117	&	-16.139	&	36.021	&	108.31	&	67.635	\\
220	&	310	&	0.15021	&	-16.135	&	36.428	&	113.91	&	67.431	\\
220	&	320	&	0.14922	&	-16.130	&	36.887	&	119.85	&	67.302	\\
220	&	330	&	0.14832	&	-16.126	&	37.374	&	125.99	&	67.164	\\
220	&	340	&	0.14746	&	-16.124	&	37.898	&	132.40	&	67.097	\\
220	&	350	&	0.14659	&	-16.121	&	38.459	&	139.13	&	67.042	\\
220	&	360	&	0.14578	&	-16.119	&	39.065	&	146.16	&	66.985	\\
230	&	180	&	0.16922	&	-16.251	&	32.199	&	49.640	&	71.317	\\
230	&	190	&	0.16696	&	-16.237	&	32.397	&	53.432	&	70.993	\\
230	&	200	&	0.16493	&	-16.225	&	32.613	&	57.315	&	70.655	\\
230	&	210	&	0.16306	&	-16.213	&	32.835	&	61.287	&	70.302	\\
230	&	220	&	0.16131	&	-16.201	&	33.063	&	65.351	&	69.938	\\
230	&	230	&	0.15975	&	-16.190	&	33.310	&	69.507	&	69.566	\\
230	&	240	&	0.15823	&	-16.178	&	33.564	&	73.778	&	69.197	\\
230	&	250	&	0.15687	&	-16.169	&	33.834	&	78.145	&	68.845	\\
230	&	260	&	0.15562	&	-16.161	&	34.153	&	82.698	&	68.509	\\
230	&	270	&	0.15443	&	-16.153	&	34.396	&	87.154	&	68.209	\\
230	&	280	&	0.15312	&	-16.145	&	34.776	&	92.162	&	67.919	\\
230	&	290	&	0.15231	&	-16.140	&	35.057	&	96.741	&	67.650	\\
230	&	300	&	0.15129	&	-16.135	&	35.435	&	101.83	&	67.408	\\
230	&	310	&	0.15029	&	-16.130	&	35.817	&	107.07	&	67.208	\\
230	&	320	&	0.14944	&	-16.126	&	36.222	&	112.41	&	67.003	\\
230	&	330	&	0.14856	&	-16.123	&	36.648	&	117.98	&	66.863	\\
230	&	340	&	0.14771	&	-16.120	&	37.104	&	123.78	&	66.747	\\
230	&	350	&	0.14691	&	-16.117	&	37.574	&	129.74	&	66.633	\\
230	&	360	&	0.14607	&	-16.114	&	38.109	&	136.12	&	66.582	\\
250	&	180	&	0.16919	&	-16.246	&	31.807	&	45.118	&	71.213	\\
250	&	190	&	0.16697	&	-16.233	&	31.979	&	48.518	&	70.865	\\
250	&	200	&	0.16492	&	-16.219	&	32.161	&	52.001	&	70.498	\\
250	&	210	&	0.16311	&	-16.208	&	32.348	&	55.532	&	70.118	\\
250	&	220	&	0.16139	&	-16.195	&	32.539	&	59.142	&	69.721	\\
250	&	230	&	0.15981	&	-16.184	&	32.748	&	62.842	&	69.324	\\
250	&	240	&	0.15840	&	-16.174	&	32.962	&	66.591	&	68.932	\\
250	&	250	&	0.15707	&	-16.164	&	33.190	&	70.435	&	68.555	\\
250	&	260	&	0.15584	&	-16.156	&	33.428	&	74.362	&	68.191	\\
250	&	270	&	0.15466	&	-16.148	&	33.682	&	78.401	&	67.858	\\
250	&	280	&	0.15352	&	-16.140	&	33.944	&	82.547	&	67.561	\\
250	&	290	&	0.15253	&	-16.134	&	34.169	&	86.620	&	67.282	\\
250	&	300	&	0.15163	&	-16.130	&	34.438	&	90.844	&	67.023	\\
250	&	310	&	0.15067	&	-16.126	&	34.828	&	95.541	&	66.772	\\
250	&	320	&	0.14981	&	-16.122	&	35.150	&	100.11	&	66.554	\\
250	&	330	&	0.14896	&	-16.117	&	35.492	&	104.83	&	66.352	\\
250	&	340	&	0.14814	&	-16.115	&	35.862	&	109.74	&	66.204	\\
250	&	350	&	0.14738	&	-16.112	&	36.237	&	114.74	&	66.047	\\
250	&	360	&	0.14660	&	-16.108	&	36.638	&	119.96	&	65.929	\\
270	&	180	&	0.16916	&	-16.243	&	31.491	&	41.370	&	71.146	\\
270	&	190	&	0.16695	&	-16.229	&	31.637	&	44.452	&	70.765	\\
270	&	200	&	0.16491	&	-16.215	&	31.788	&	47.595	&	70.374	\\
270	&	210	&	0.16309	&	-16.202	&	31.959	&	50.804	&	69.964	\\
270	&	220	&	0.16141	&	-16.190	&	32.114	&	54.040	&	69.544	\\
270	&	230	&	0.15988	&	-16.179	&	32.290	&	57.348	&	69.125	\\
270	&	240	&	0.15849	&	-16.169	&	32.473	&	60.706	&	68.717	\\
270	&	250	&	0.15718	&	-16.160	&	32.672	&	64.155	&	68.323	\\
270	&	260	&	0.15596	&	-16.151	&	32.875	&	67.661	&	67.944	\\
270	&	270	&	0.15485	&	-16.144	&	33.091	&	71.235	&	67.582	\\
270	&	280	&	0.15373	&	-16.136	&	33.299	&	74.877	&	67.274	\\
270	&	290	&	0.15275	&	-16.131	&	33.554	&	78.645	&	66.958	\\
270	&	300	&	0.15173	&	-16.125	&	33.747	&	82.375	&	66.677	\\
270	&	310	&	0.15099	&	-16.121	&	34.025	&	86.245	&	66.435	\\
270	&	320	&	0.15012	&	-16.118	&	34.259	&	90.157	&	66.176	\\
270	&	330	&	0.14928	&	-16.114	&	34.603	&	94.438	&	65.981	\\
270	&	340	&	0.14850	&	-16.110	&	34.906	&	98.668	&	65.774	\\
270	&	350	&	0.14776	&	-16.107	&	35.219	&	102.99	&	65.578	\\
270	&	360	&	0.14698	&	-16.103	&	35.565	&	107.55	&	65.419	\\
300	&	180	&	0.16903	&	-16.236	&	31.097	&	36.794	&	71.044	\\
300	&	190	&	0.16689	&	-16.223	&	31.220	&	39.492	&	70.646	\\
300	&	200	&	0.16493	&	-16.210	&	31.345	&	42.234	&	70.223	\\
300	&	210	&	0.16314	&	-16.197	&	31.485	&	45.031	&	69.784	\\
300	&	220	&	0.16149	&	-16.185	&	31.613	&	47.853	&	69.336	\\
300	&	230	&	0.15998	&	-16.174	&	31.755	&	50.725	&	68.894	\\
300	&	240	&	0.15857	&	-16.163	&	31.903	&	53.649	&	68.463	\\
300	&	250	&	0.15730	&	-16.154	&	32.066	&	56.624	&	68.049	\\
300	&	260	&	0.15613	&	-16.146	&	32.231	&	59.638	&	67.659	\\
300	&	270	&	0.15501	&	-16.138	&	32.403	&	62.712	&	67.284	\\
300	&	280	&	0.15395	&	-16.131	&	32.584	&	65.847	&	66.930	\\
300	&	290	&	0.15297	&	-16.125	&	32.771	&	69.031	&	66.626	\\
300	&	300	&	0.15207	&	-16.120	&	32.970	&	72.269	&	66.318	\\
300	&	310	&	0.15130	&	-16.118	&	33.194	&	75.570	&	65.986	\\
300	&	320	&	0.15034	&	-16.112	&	33.404	&	79.002	&	65.748	\\
300	&	330	&	0.14964	&	-16.108	&	33.607	&	82.346	&	65.479	\\
300	&	340	&	0.14884	&	-16.104	&	33.845	&	85.904	&	65.259	\\
300	&	350	&	0.14812	&	-16.100	&	34.084	&	89.488	&	65.077	\\
300	&	360	&	0.14739	&	-16.096	&	34.335	&	93.180	&	64.860	\\
350	&	180	&	0.16900	&	-16.231	&	30.624	&	31.064	&	70.942	\\
350	&	190	&	0.16688	&	-16.217	&	30.711	&	33.301	&	70.490	\\
350	&	200	&	0.16492	&	-16.203	&	30.803	&	35.575	&	70.026	\\
350	&	210	&	0.16312	&	-16.189	&	30.907	&	37.894	&	69.550	\\
350	&	220	&	0.16149	&	-16.177	&	31.006	&	40.228	&	69.072	\\
350	&	230	&	0.16005	&	-16.166	&	31.107	&	42.574	&	68.596	\\
350	&	240	&	0.15867	&	-16.155	&	31.220	&	44.975	&	68.134	\\
350	&	250	&	0.15743	&	-16.146	&	31.346	&	47.408	&	67.709	\\
350	&	260	&	0.15629	&	-16.139	&	31.463	&	49.849	&	67.298	\\
350	&	270	&	0.15518	&	-16.130	&	31.593	&	52.352	&	66.918	\\
350	&	280	&	0.15416	&	-16.123	&	31.718	&	54.866	&	66.566	\\
350	&	290	&	0.15322	&	-16.118	&	31.862	&	57.433	&	66.208	\\
350	&	300	&	0.15232	&	-16.112	&	32.008	&	60.039	&	65.868	\\
350	&	310	&	0.15150	&	-16.107	&	32.159	&	62.670	&	65.566	\\
350	&	320	&	0.15071	&	-16.102	&	32.314	&	65.342	&	65.241	\\
350	&	330	&	0.14994	&	-16.098	&	32.478	&	68.077	&	64.982	\\
350	&	340	&	0.14923	&	-16.095	&	32.653	&	70.853	&	64.737	\\
350	&	350	&	0.14850	&	-16.091	&	32.843	&	73.720	&	64.484	\\
350	&	360	&	0.14788	&	-16.087	&	33.004	&	76.518	&	64.176	\\
400	&	180	&	0.16900	&	-16.227	&	30.281	&	26.877	&	70.843	\\
400	&	190	&	0.16684	&	-16.212	&	30.347	&	28.799	&	70.370	\\
400	&	200	&	0.16492	&	-16.198	&	30.422	&	30.744	&	69.878	\\
400	&	210	&	0.16316	&	-16.184	&	30.498	&	32.712	&	69.364	\\
400	&	220	&	0.16154	&	-16.171	&	30.572	&	34.698	&	68.846	\\
400	&	230	&	0.16007	&	-16.159	&	30.665	&	36.718	&	68.364	\\
400	&	240	&	0.15851	&	-16.145	&	30.729	&	38.774	&	68.006	\\
400	&	250	&	0.15747	&	-16.139	&	30.833	&	40.792	&	67.481	\\
400	&	260	&	0.15627	&	-16.129	&	30.919	&	42.870	&	67.040	\\
400	&	270	&	0.15529	&	-16.125	&	31.032	&	44.963	&	66.660	\\
400	&	280	&	0.15427	&	-16.117	&	31.136	&	47.093	&	66.260	\\
400	&	290	&	0.15334	&	-16.111	&	31.231	&	49.221	&	65.925	\\
400	&	300	&	0.15248	&	-16.105	&	31.360	&	51.416	&	65.556	\\
400	&	310	&	0.15166	&	-16.101	&	31.467	&	53.601	&	65.236	\\
400	&	320	&	0.15088	&	-16.095	&	31.592	&	55.837	&	64.908	\\
400	&	330	&	0.15012	&	-16.090	&	31.715	&	58.097	&	64.593	\\
400	&	340	&	0.14946	&	-16.088	&	31.862	&	60.400	&	64.309	\\
400	&	350	&	0.14882	&	-16.085	&	31.982	&	62.678	&	64.013	\\
400	&	360	&	0.14821	&	-16.083	&	32.131	&	65.037	&	63.769	\\
500	&	180	&	0.16897	&	-16.222	&	29.827	&	21.182	&	70.716	\\
500	&	190	&	0.16685	&	-16.206	&	29.869	&	22.676	&	70.193	\\
500	&	200	&	0.16489	&	-16.190	&	29.911	&	24.186	&	69.664	\\
500	&	210	&	0.16310	&	-16.174	&	29.952	&	25.710	&	69.091	\\
500	&	220	&	0.16153	&	-16.160	&	30.000	&	27.240	&	68.518	\\
500	&	230	&	0.16009	&	-16.150	&	30.055	&	28.787	&	68.054	\\
500	&	240	&	0.15879	&	-16.141	&	30.114	&	30.344	&	67.607	\\
500	&	250	&	0.15754	&	-16.130	&	30.174	&	31.921	&	67.142	\\
500	&	260	&	0.15625	&	-16.118	&	30.230	&	33.534	&	66.756	\\
500	&	270	&	0.15542	&	-16.116	&	30.310	&	35.103	&	66.268	\\
500	&	280	&	0.15444	&	-16.108	&	30.376	&	36.716	&	65.885	\\
500	&	290	&	0.15348	&	-16.101	&	30.445	&	38.352	&	65.520	\\
500	&	300	&	0.15264	&	-16.096	&	30.523	&	39.994	&	65.155	\\
500	&	310	&	0.15184	&	-16.090	&	30.601	&	41.652	&	64.764	\\
500	&	320	&	0.15111	&	-16.085	&	30.680	&	43.313	&	64.419	\\
500	&	330	&	0.15039	&	-16.080	&	30.762	&	45.000	&	64.076	\\
500	&	340	&	0.14973	&	-16.077	&	30.849	&	46.699	&	63.761	\\
500	&	350	&	0.14917	&	-16.075	&	30.943	&	48.401	&	63.475	\\
500	&	360	&	0.14839	&	-16.068	&	31.023	&	50.175	&	63.242	\\
600	&	180	&	0.16891	&	-16.216	&	29.535	&	17.486	&	70.604	\\
600	&	190	&	0.16679	&	-16.200	&	29.559	&	18.706	&	70.056	\\
600	&	200	&	0.16486	&	-16.184	&	29.586	&	19.940	&	69.507	\\
600	&	210	&	0.16311	&	-16.169	&	29.611	&	21.179	&	68.924	\\
600	&	220	&	0.16154	&	-16.155	&	29.642	&	22.428	&	68.355	\\
600	&	230	&	0.16000	&	-16.141	&	29.678	&	23.702	&	67.871	\\
600	&	240	&	0.15879	&	-16.134	&	29.719	&	24.954	&	67.402	\\
600	&	250	&	0.15758	&	-16.125	&	29.765	&	26.235	&	66.936	\\
600	&	260	&	0.15651	&	-16.116	&	29.807	&	27.509	&	66.482	\\
600	&	270	&	0.15543	&	-16.108	&	29.856	&	28.813	&	66.047	\\
600	&	280	&	0.15446	&	-16.101	&	29.905	&	30.116	&	65.627	\\
600	&	290	&	0.15355	&	-16.094	&	29.956	&	31.430	&	65.245	\\
600	&	300	&	0.15271	&	-16.089	&	30.012	&	32.755	&	64.884	\\
600	&	310	&	0.15189	&	-16.083	&	30.067	&	34.091	&	64.527	\\
600	&	320	&	0.15122	&	-16.079	&	30.130	&	35.421	&	64.151	\\
600	&	330	&	0.15056	&	-16.075	&	30.184	&	36.754	&	63.778	\\
600	&	340	&	0.14971	&	-16.068	&	30.242	&	38.157	&	63.578	\\
600	&	350	&	0.14911	&	-16.064	&	30.310	&	39.525	&	63.238	\\
600	&	360	&	0.14843	&	-16.057	&	30.365	&	40.916	&	62.903	\\
800	&	180	&	0.16881	&	-16.209	&	29.179	&	12.964	&	70.422	\\
800	&	190	&	0.16671	&	-16.192	&	29.185	&	13.859	&	69.846	\\
800	&	200	&	0.16480	&	-16.176	&	29.193	&	14.762	&	69.272	\\
800	&	210	&	0.16304	&	-16.160	&	29.199	&	15.671	&	68.706	\\
800	&	220	&	0.16150	&	-16.147	&	29.212	&	16.580	&	68.158	\\
800	&	230	&	0.16003	&	-16.135	&	29.229	&	17.504	&	67.664	\\
800	&	240	&	0.15876	&	-16.125	&	29.247	&	18.422	&	67.160	\\
800	&	250	&	0.15752	&	-16.114	&	29.270	&	19.357	&	66.693	\\
800	&	260	&	0.15641	&	-16.106	&	29.296	&	20.291	&	66.231	\\
800	&	270	&	0.15547	&	-16.099	&	29.317	&	21.214	&	65.771	\\
800	&	280	&	0.15450	&	-16.092	&	29.365	&	22.174	&	65.337	\\
800	&	290	&	0.15366	&	-16.087	&	29.389	&	23.110	&	64.926	\\
800	&	300	&	0.15284	&	-16.081	&	29.423	&	24.064	&	64.539	\\
800	&	310	&	0.15209	&	-16.076	&	29.463	&	25.022	&	64.132	\\
800	&	320	&	0.15137	&	-16.071	&	29.500	&	25.985	&	63.781	\\
800	&	330	&	0.15067	&	-16.066	&	29.532	&	26.950	&	63.423	\\
800	&	340	&	0.15002	&	-16.061	&	29.569	&	27.922	&	63.065	\\
800	&	350	&	0.14945	&	-16.058	&	29.612	&	28.894	&	62.718	\\
800	&	360	&	0.14893	&	-16.056	&	29.658	&	29.871	&	62.405	\\
1000	&	180	&	0.16872	&	-16.202	&	28.970	&	10.302	&	70.266	\\
1000	&	190	&	0.16664	&	-16.185	&	28.964	&	11.008	&	69.677	\\
1000	&	200	&	0.16476	&	-16.169	&	28.962	&	11.719	&	69.090	\\
1000	&	210	&	0.16299	&	-16.153	&	28.960	&	12.437	&	68.513	\\
1000	&	220	&	0.16142	&	-16.139	&	28.959	&	13.156	&	67.962	\\
1000	&	230	&	0.16000	&	-16.128	&	28.971	&	13.882	&	67.447	\\
1000	&	240	&	0.15871	&	-16.118	&	28.979	&	14.608	&	66.975	\\
1000	&	250	&	0.15747	&	-16.107	&	28.989	&	15.341	&	66.512	\\
1000	&	260	&	0.15643	&	-16.100	&	29.011	&	16.072	&	66.025	\\
1000	&	270	&	0.15538	&	-16.092	&	29.022	&	16.810	&	65.611	\\
1000	&	280	&	0.15445	&	-16.085	&	29.040	&	17.549	&	65.178	\\
1000	&	290	&	0.15367	&	-16.080	&	29.063	&	18.282	&	64.717	\\
1000	&	300	&	0.15288	&	-16.075	&	29.088	&	19.027	&	64.325	\\
1000	&	310	&	0.15209	&	-16.069	&	29.113	&	19.780	&	63.931	\\
1000	&	320	&	0.15135	&	-16.063	&	29.135	&	20.533	&	63.590	\\
1000	&	330	&	0.15079	&	-16.060	&	29.167	&	21.276	&	63.155	\\
1000	&	340	&	0.15011	&	-16.056	&	29.191	&	22.040	&	62.850	\\
1000	&	350	&	0.14952	&	-16.052	&	29.220	&	22.801	&	62.501	\\
1000	&	360	&	0.14896	&	-16.048	&	29.247	&	23.561	&	62.158	\\
1200	&	180	&	0.16875	&	-16.199	&	28.835	&	8.5436	&	70.139	\\
1200	&	190	&	0.16663	&	-16.181	&	28.825	&	9.1298	&	69.544	\\
1200	&	200	&	0.16460	&	-16.163	&	28.822	&	9.7278	&	68.962	\\
1200	&	210	&	0.16296	&	-16.149	&	28.802	&	10.310	&	68.395	\\
1200	&	220	&	0.16138	&	-16.135	&	28.801	&	10.907	&	67.842	\\
1200	&	230	&	0.15997	&	-16.123	&	28.799	&	11.502	&	67.322	\\
1200	&	240	&	0.15865	&	-16.111	&	28.800	&	12.102	&	66.812	\\
1200	&	250	&	0.15749	&	-16.103	&	28.810	&	12.704	&	66.354	\\
1200	&	260	&	0.15639	&	-16.095	&	28.822	&	13.310	&	65.891	\\
1200	&	270	&	0.15538	&	-16.087	&	28.827	&	13.914	&	65.443	\\
1200	&	280	&	0.15449	&	-16.081	&	28.844	&	14.521	&	65.010	\\
1200	&	290	&	0.15364	&	-16.075	&	28.855	&	15.129	&	64.584	\\
1200	&	300	&	0.15286	&	-16.069	&	28.876	&	15.741	&	64.156	\\
1200	&	310	&	0.15211	&	-16.064	&	28.894	&	16.357	&	63.770	\\
1200	&	320	&	0.15140	&	-16.059	&	28.908	&	16.973	&	63.399	\\
1200	&	330	&	0.15075	&	-16.055	&	28.927	&	17.590	&	63.031	\\
1200	&	340	&	0.15011	&	-16.050	&	28.948	&	18.213	&	62.684	\\
1200	&	350	&	0.14954	&	-16.046	&	28.969	&	18.834	&	62.321	\\
1200	&	360	&	0.14898	&	-16.042	&	28.984	&	19.456	&	61.961	\\
1400	&	180	&	0.16865	&	-16.194	&	28.738	&	7.3027	&	70.034	\\
1400	&	190	&	0.16657	&	-16.177	&	28.721	&	7.8002	&	69.442	\\
1400	&	200	&	0.16466	&	-16.160	&	28.709	&	8.3026	&	68.863	\\
1400	&	210	&	0.16291	&	-16.144	&	28.692	&	8.8063	&	68.303	\\
1400	&	220	&	0.16137	&	-16.132	&	28.689	&	9.3124	&	67.758	\\
1400	&	230	&	0.15990	&	-16.118	&	28.679	&	9.8218	&	67.240	\\
1400	&	240	&	0.15861	&	-16.109	&	28.683	&	10.333	&	66.762	\\
1400	&	250	&	0.15747	&	-16.100	&	28.682	&	10.842	&	66.276	\\
1400	&	260	&	0.15638	&	-16.091	&	28.688	&	11.356	&	65.788	\\
1400	&	270	&	0.15538	&	-16.083	&	28.692	&	11.871	&	65.315	\\
1400	&	280	&	0.15444	&	-16.076	&	28.699	&	12.388	&	64.893	\\
1400	&	290	&	0.15359	&	-16.069	&	28.712	&	12.908	&	64.472	\\
1400	&	300	&	0.15282	&	-16.064	&	28.720	&	13.424	&	64.049	\\
1400	&	310	&	0.15208	&	-16.060	&	28.739	&	13.948	&	63.635	\\
1400	&	320	&	0.15142	&	-16.055	&	28.749	&	14.465	&	63.261	\\
1400	&	330	&	0.15075	&	-16.050	&	28.765	&	14.992	&	62.879	\\
1400	&	340	&	0.15013	&	-16.046	&	28.776	&	15.516	&	62.524	\\
1400	&	350	&	0.14955	&	-16.042	&	28.794	&	16.045	&	62.178	\\
1400	&	360	&	0.14898	&	-16.038	&	28.808	&	16.574	&	61.836	\\
1800	&	180	&	0.16864	&	-16.189	&	28.611	&	5.6552	&	69.856	\\
1800	&	190	&	0.16651	&	-16.170	&	28.588	&	6.0409	&	69.269	\\
1800	&	200	&	0.16459	&	-16.154	&	28.568	&	6.4285	&	68.698	\\
1800	&	210	&	0.16289	&	-16.138	&	28.549	&	6.8159	&	68.148	\\
1800	&	220	&	0.16128	&	-16.126	&	28.538	&	7.2087	&	67.634	\\
1800	&	230	&	0.15994	&	-16.114	&	28.518	&	7.5943	&	67.105	\\
1800	&	240	&	0.15860	&	-16.104	&	28.523	&	7.9932	&	66.608	\\
1800	&	250	&	0.15737	&	-16.093	&	28.516	&	8.3890	&	66.118	\\
1800	&	260	&	0.15627	&	-16.083	&	28.511	&	8.7843	&	65.653	\\
1800	&	270	&	0.15530	&	-16.076	&	28.510	&	9.1788	&	65.171	\\
1800	&	280	&	0.15437	&	-16.069	&	28.515	&	9.5780	&	64.756	\\
1800	&	290	&	0.15352	&	-16.063	&	28.518	&	9.9763	&	64.327	\\
1800	&	300	&	0.15273	&	-16.057	&	28.522	&	10.375	&	63.912	\\
1800	&	310	&	0.15204	&	-16.052	&	28.528	&	10.772	&	63.475	\\
1800	&	320	&	0.15138	&	-16.048	&	28.543	&	11.173	&	63.066	\\
1800	&	330	&	0.15074	&	-16.043	&	28.549	&	11.574	&	62.659	\\
1800	&	340	&	0.15012	&	-16.039	&	28.549	&	11.974	&	62.326	\\
1800	&	350	&	0.14953	&	-16.035	&	28.566	&	12.382	&	61.997	\\
1800	&	360	&	0.14896	&	-16.031	&	28.575	&	12.789	&	61.660	\\
\hline
\end{longtable}%

\end{document}